\pdfoutput=1
\documentclass[11pt,twoside,a4paper,cmspaper,final,collab]{cms-tdr}
\def\svnVersion{d50db3b}\def\svnDate{2024/07/30}\def\cmsCernNoTag{CERN-EP-2024-066}\def\cmsCernDate{\today}\def\cmsMessage{Published in the European Physical Journal C as \href{http://dx.doi.org/10.1140/epjc/s10052-024-13116-7}{\doi{10.1140/epjc/s10052-024-13116-7}.}}
\begin{document}\cmsNoteHeader{SMP-22-005}

\providecommand{\alpsmz}{\ensuremath{\alpS(m_{\PZ})}\xspace}
\newcommand{\alpsq}{\ensuremath{\alpS(Q)}\xspace}
\newcommand{\alpsquared}{\ensuremath{\alpS^{2}}\xspace}
\newcommand{\alpscubed}{\ensuremath{\alpS^{3}}\xspace}
\newcommand{\chisq}{\ensuremath{\chi^2}\xspace}
\newcommand{\chisqndof}{\ensuremath{\chi^2/n_\text{dof}}\xspace}
\newcommand{\NLOJETPP} {{\textsc{NLOJet++}}\xspace}
\providecommand{\fastNLO} {{\textsc{fastNLO}}\xspace}
\providecommand{\PYTHIAEIGHT} {{\textsc{pythia8}}\xspace}
\newcommand{\hatHTtwo}{\ensuremath{\hat{H}_{\mathrm{T}}/2}\xspace}
\newcommand{\hatHT}{\ensuremath{\hat{H}_{\mathrm{T}}}\xspace}
\newcommand{\mur}{\ensuremath{\mu_\mathrm{R}}\xspace}
\newcommand{\muf}{\ensuremath{\mu_\mathrm{F}}\xspace}
\newcommand{\yabs}{\ensuremath{\abs{y}}\xspace}
\newcommand{\rdphi}{\ensuremath{R_{\Delta\phi}(\pt)}\xspace}
\newcommand{\pTnbrmin}{\ensuremath{p_{\text{Tmin}}^{\text{nbr}}}\xspace}
\newcommand{\Nptn}{\ensuremath{N(\pt,n)}\xspace}
\providecommand{\pbinvns}{\mbox{\ensuremath{\text{pb}^{-1}}}\xspace}

\newlength\cmsFigWidthA
\ifthenelse{\boolean{cms@external}}{\setlength\cmsFigWidthA{0.47\textwidth}}{\setlength\cmsFigWidthA{0.97\textwidth}}

\cmsNoteHeader{SMP-22-005}
\title{Measurement of multijet azimuthal correlations and determination of the strong coupling in proton-proton collisions at \texorpdfstring{$\sqrt{s}=13\TeV$}{sqrt(s)=13 TeV}}

\date{\today}

\abstract{{\tolerance=800 A measurement is presented of a ratio observable that provides a measure of the azimuthal correlations among jets with large transverse momentum \pt. This observable is measured in multijet events over the range of $\pt = 360$--$3170\GeV$ based on data collected by the CMS experiment in proton-proton collisions at a centre-of-mass energy of 13\TeV, corresponding to an integrated luminosity of 134\fbinv. The results are compared with predictions from Monte Carlo parton-shower event generator simulations, as well as with fixed-order perturbative quantum chromodynamics (pQCD) predictions at next-to-leading-order (NLO) accuracy obtained with different parton distribution functions (PDFs) and corrected for nonperturbative and electroweak effects. Data and theory agree within uncertainties. From the comparison of the measured observable with the pQCD prediction obtained with the NNPDF3.1 NLO PDFs, the strong coupling at the \PZ boson mass scale is $\alpsmz=0.1177 \pm 0.0013\, \text{(exp)} _{-0.0073}^{+0.0116} \thy = 0.1177_{-0.0074}^{+0.0117}$, where the total uncertainty is dominated by the scale dependence of the fixed-order predictions. A test of the running of \alpS in the \TeV region shows no deviation from the expected NLO pQCD behaviour.\par}}

\hypersetup{
pdfauthor={CMS Collaboration},
pdftitle={Measurement of multijet azimuthal correlations and determination of the strong coupling in proton-proton collisions at sqrt(s) = 13 TeV},
pdfsubject={CMS},
pdfkeywords={CMS, QCD, jets, azimuthal angle, MC generators, PDFs, strong coupling}}

\maketitle 

\section{Introduction}
\label{sec:intro}

In the standard model of particle physics, the strong interaction between partons (quarks and gluons) is described by the theory of quantum chromodynamics (QCD). A key property of the strong interaction is ``asymptotic freedom'', which characterizes the decreasing value of the coupling \alpsq for increasingly larger momentum transfer $Q$ that corresponds to smaller distances between the interacting partons. This property is a consequence of the non-Abelian nature of QCD, and can be theoretically derived from the renormalization group equations (RGE)~\cite{Callan:1970yg,Symanzik:1970rt,Symanzik:1971vw}. Although the RGE cannot predict the absolute value of \alpsq, they can accurately determine its evolution as a function of the energy scale $Q$~\cite{Baikov:2016tgj}. By comparing experimental measurements to perturbative QCD (pQCD) predictions for a given observable, the value of \alpsq can be extracted at various scales~\cite{PDG,dEnterria:2022hzv}. To compare various \alpsq determinations, it is standard practice to evolve them to a common scale given by the mass of the \PZ boson, $Q = m_{\PZ}$. The current world-average value of the QCD coupling at this reference scale is $\alpsmz = 0.1180 \pm 0.0009$~\cite{PDG}.

\begin{figure*}[hbt!]
  \centering
	\includegraphics[width=0.98\textwidth]{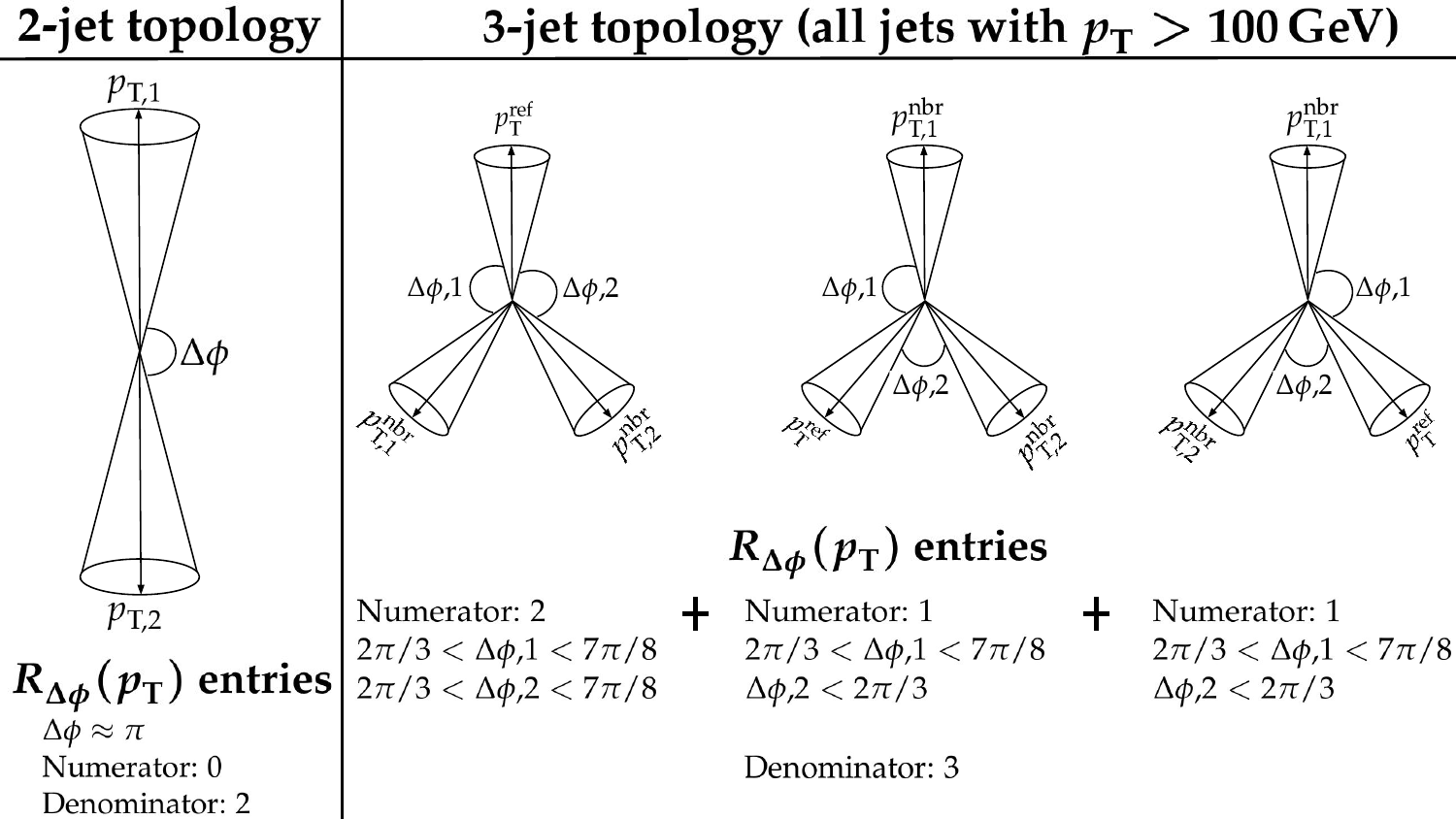}
	\caption{Example of the number of entries contributing to the numerator and denominator of the \rdphi ratio, Eq.~(\ref{eq:observable}), for 2-jet (left) and 3-jet (right) events, with all jets having $\pt > \pTnbrmin=100\GeV$. The 2-jet topology does not contribute (null numerator) to the \rdphi ratio when the azimuthal distance for neighbouring jets is fixed to $2\pi/3<\Delta\phi<7\pi/8$. In the 3-jet topology, each jet is considered as a reference, and its azimuthal separations ($\Delta\phi{,}1$ and $\Delta\phi{,}2$) to other neighbouring jets (with $p_{\mathrm {T,1}}^{\text {nbr}}$ and $p_{\mathrm{T,2}}^{\text {nbr}}$) are computed. Each neighbouring jet with $\Delta\phi$ within the specified interval increments the entries of the numerator, whereas the denominator simply counts the number of jets in the event.}
\label{fig:Sketch}
\end{figure*}

This paper reports a new extraction of the \alpsq coupling from multijet measurements at various energy scales in proton-proton ($\Pp\Pp$) collisions at the CERN LHC. For this purpose, a ratio observable \rdphi, related to the azimuthal correlations among jets, is measured as a function of the jet transverse momentum \pt. Similar ratio observables, based on either the distance in the plane of rapidity and azimuthal angle among jets, $R_{\Delta R}(\pt)$~\cite{bib:D02}, or on the dijet azimuthal decorrelations, $R_{\Delta\phi}(H_\mathrm{T})$~\cite{Wobisch:2012au,bib:ATLAS8TeVazi}, have already been used to extract the \alpsq coupling at hadron colliders. The \rdphi observable is defined as:
\begin{equation}
  \rdphi =
  \frac{\sum_{i=1}^{N_{\text{jet}}(\pt)} N_{\text{nbr}}^{(i)}(\Delta\phi,\pTnbrmin)}
  {N_{\text{jet}}(\pt)},
   \label{eq:observable}
\end{equation}
where the denominator $N_{\text{jet}}(\pt)$ simply counts the number of jets in a given jet \pt bin, and the numerator sums the number of neighbouring jets, $N_{\text{nbr}}^{(i)}$, around each jet $i$ in the same \pt bin.
A neighbouring jet must exceed a minimum transverse momentum of \pTnbrmin and be separated from jet $i$ within a specified interval of azimuthal distance $\Delta\phi$:~${\Delta\phi}_{\text{min}}<\Delta\phi<{\Delta\phi}_{\text{max}}$.
In fixed-order predictions of jet production based on pQCD calculations, the leading-order (LO) $2\to 2$ process is characterized by an azimuthal separation of $\Delta\phi = \pi$.
Since the sum in the numerator runs over all jets, this would lead to two entries at $\pt = p_\mathrm{T,1} = p_\mathrm{T,2}$.
At next-to-leading order (NLO), the radiation of a third hard parton can give rise to a 3-jet topology with $\Delta\phi$ between $2\pi/3$ and $\pi$ with respect to the jet opposite to the hemisphere with radiation (Fig.~\ref{fig:Sketch}, right diagrams). 
Hence, by fixing the azimuthal distance for neighbouring jets to $2\pi/3<\Delta\phi<7\pi/8$ in Eq.~(\ref{eq:observable}), the dijet case is avoided, and the numerator is different from zero only for events with three jets or more, whose LO cross section is proportional to \alpscubed.
Also here, a jet pair fulfilling both the selection in \pTnbrmin and in $\Delta\phi$ leads to two entries, but potentially at different jet \pt values.
On the other hand, the denominator corresponds to the inclusive jet cross section, which at LO is proportional to \alpsquared, such that the \rdphi observable is directly proportional to \alpS , at the lowest order.
A representative illustration, indicating the entries to the numerator and denominator of the \rdphi ratio, is shown in the left and right panels of Fig.~\ref{fig:Sketch} for a 2-jet and a 3-jet event, respectively. 

In the ratio defined by Eq.~(\ref{eq:observable}) many experimental systematic uncertainties---such as those from the integrated luminosity, the jet energy scale (JES), and the jet energy resolution (JER)---cancel entirely or to a large extent.
In addition, theoretical uncertainties---such as nonperturbative (NP) and parton distribution function (PDF) uncertainties---are reduced.

To rigorously account for correlations between the numerator and denominator, it is useful to consider the more general, two-dimensional jet-counting quantity, \Nptn, which is a function of the $i$-th jet's \pt and of the number $n$ of neighbouring jets that satisfy the additional selection criteria for \pTnbrmin and $\Delta\phi$.
Then, using \Nptn, it can be shown that the \rdphi observable can be also formulated as:
\begin{equation}
  \label{eq:observableEquiv}
  \rdphi = \frac{\sum_{n}n\Nptn}{\sum_{n}\Nptn}.
\end{equation}
Such a definition allows a multidimensional unfolding of the more general quantity \Nptn to be performed, instead of a separate unfolding of the numerator and denominator of Eq.~(\ref{eq:observable}).

The measurement is performed using data collected with the CMS detector, during the LHC Run~2 data-taking period (2016--2018), corresponding to an integrated luminosity of 134\fbinv at a centre-of-mass energy of 13\TeV~\cite{CMS-LUM-17-003,CMS-PAS-LUM-17-004,CMS-PAS-LUM-18-002}.
Previous determinations of the strong coupling constant \alpsmz using jets at hadron colliders have been reported by the CDF~\cite{bib:CDF} and D0~\cite{bib:D01,bib:D02} Collaborations in proton-antiproton collisions at $\sqrt{s}=1.8$ and 1.96\TeV at the Fermilab Tevatron.
At the LHC, determinations have been reported using $\Pp\Pp$ collision data from the ATLAS and CMS Collaborations at $\sqrt{s}=7$~\cite{bib:ATLAS1,bib:R32,bib:ATLAS7TeV,bib:TOP,bib:3jetmass,bib:7TeVas,CMS:2019oeb,dEnterria:2019aat}, 8~\cite{bib:ATLAS8TeVTEEC,bib:8TeVas,bib:CMS8TeV3D,bib:ATLAS8TeVazi,CMS:2019oeb,dEnterria:2019aat}, and 13~\cite{CMS:2018ypj,bib:TOP013,bib:TOP113,bib:ATLAS13TeV,CMS:2021yzl,bib:ATLAS13TeV, CMS:2023fix,CMS:2024mlf}\TeV. 

The paper is organized as follows.
In Section~\ref{sec:detector} a brief description of the CMS detector is given. 
In Section~\ref{sec:Reconstruction} the event reconstruction is described. Section~\ref{sec:Measurement} details the measurement of the \rdphi observable.
Experimental results and theoretical predictions for the \rdphi observable are compared in Section~\ref{sec:TheoreticalPredictions}.
The determination of \alpsmz and the investigation of the running of the \alpsq coupling are presented in Section~\ref{sec:AlphaS}.
Finally, a summary of the paper is given in Section~\ref{sec:summary}. 

{\tolerance=800
Tabulated results are provided in the HEPData record for this analysis~\cite{hepdata}.
\par}

\section{The CMS detector}
\label{sec:detector}

The central feature of the CMS apparatus is a superconducting solenoid of 6\unit{m} internal diameter, providing a magnetic field of 3.8\unit{T}. Within the solenoid volume are a silicon pixel and strip tracker, a lead tungstate crystal electromagnetic calorimeter (ECAL), and a brass and scintillator hadron calorimeter (HCAL), each composed of a barrel and two endcap sections. Forward calorimeters extend the pseudorapidity coverage provided by the barrel and endcap detectors. Muons are measured in gas-ionization detectors embedded in the steel flux-return yoke outside the solenoid.

The electromagnetic calorimeter consists of 75\,848 lead tungstate crystals, which provide coverage in pseudorapidity $\abs{\eta} < 1.48$ in a barrel region (EB) and $1.48 < \abs{\eta} < 3.0$ in two endcap regions (EE). Preshower detectors consisting of two planes of silicon sensors interleaved with a total of three~radiation lengths of lead are located in front of each EE detector. 

In the region $\abs{\eta} < 1.74$, the HCAL cells have widths of 0.087 in pseudorapidity and 0.087 in azimuth ($\phi$). In the $\eta$--$\phi$ plane, and for $\abs{\eta} < 1.48$, the HCAL cells map on to $5{\times}5$ arrays of ECAL crystals to form calorimeter towers projecting radially outwards from close to the nominal interaction point. For $\abs{\eta} > 1.74$, the coverage of the towers increases progressively to a maximum of 0.174 in $\Delta \eta$ and $\Delta \phi$. A more detailed description of the CMS detector, together with a definition of the coordinate system used and the relevant kinematic variables, can be found in Ref.~\cite{CMS:2008xjf}. 

Events of interest are selected using a two-tiered trigger system. The first level (L1), composed of custom hardware processors, uses information from the calorimeters and muon detectors to select events at a rate of around 100\unit{kHz} within a fixed latency of about 4\mus~\cite{CMS:2020cmk}. The second level, known as the high-level trigger (HLT), consists of a farm of processors running a version of the full event reconstruction software optimized for fast processing and reduces the event rate to around 1\unit{kHz} before data storage~\cite{CMS:2016ngn}.

\section{Event reconstruction}
\label{sec:Reconstruction}

The global event reconstruction---also called particle-flow (PF) event reconstruction~\cite{CMS:2017yfk}---aims to reconstruct and identify each individual particle in an event, with an optimized combination of all subdetector information.
In this process, the identification of the particle type (photon, electron, muon, charged hadron, neutral hadron) plays an important role in the determination of the particle direction and energy.
Photons (\eg, coming from \PGpz decays or from electron bremsstrahlung) are identified as ECAL energy clusters not linked to the extrapolation of any charged particle trajectory to the ECAL.
Electrons (\eg, coming from photon conversions in the tracker material or from bottom quark (\PQb) hadron semileptonic decays) are identified as a primary charged particle track and potentially many ECAL energy clusters corresponding to this track extrapolation to the ECAL and to possible bremsstrahlung photons emitted along the way through the tracker material.
Muons (\eg, from \PQb hadron semileptonic decays) are identified as tracks in the central tracker consistent with either a track or several hits in the muon system, and associated with calorimeter deposits compatible with the muon hypothesis.
Charged hadrons are identified as charged particle tracks neither identified as electrons, nor as muons.
Finally, neutral hadrons are identified as HCAL energy clusters not linked to any charged-hadron trajectory, or as a combined ECAL and HCAL energy excess with respect to the expected charged-hadron energy deposit.

The energy of photons is obtained from the ECAL measurement.
The energy of electrons is determined from a combination of the track momentum at the main interaction vertex, the corresponding ECAL cluster energy, and the energy sum of all bremsstrahlung photons attached to the track.
The energy of muons is obtained from the corresponding track momentum.
The energy of charged hadrons is determined from a combination of the track momentum and the corresponding ECAL and HCAL energies, corrected for the response function of the calorimeters to hadronic showers.
Finally, the energy of neutral hadrons is obtained from the corresponding corrected ECAL and HCAL energies.

The primary vertex (PV) is taken to be the vertex corresponding to the hardest scattering in the event, evaluated using tracking information alone, as described in Section 9.4.1 of Ref.~\cite{CMS-TDR-15-02}.
For each event, hadronic jets are clustered from the reconstructed particle candidates using the infrared and collinear safe anti-\kt algorithm~\cite{Cacciari:2008gp,Cacciari:2011ma} with a distance parameter of $R=0.7$.
This choice of the parameter $R$ enables the compatibility with previous results from the CMS Collaboration at $\sqrt{s}=7$~\cite{bib:R32} and 13~\cite{CMS:2021yzl}\TeV. 
Jet momentum is determined as the vectorial sum of all particle momenta in the jet, and is found from simulation to be, on average, within 5 to 10\% of the true momentum over the whole \pt spectrum and detector acceptance.
Additional $\Pp\Pp$ interactions within the same, or nearby, bunch crossings (pileup) can contribute additional tracks and calorimetric energy depositions to the jet momentum.
To mitigate this effect, charged particles identified to be originating from pileup vertices are discarded, and an offset correction is applied to correct for the remaining contributions~\cite{CMS:2020ebo}.
The JES corrections are derived from simulation to bring the measured response of jets to that of particle-level jets on average. In situ measurements of the momentum balance in dijet, $\text{photon} + \text{jet}$, $\PZ + \text{jet}$, and multijet events are used to account for any residual differences in the jet energy scale between the measured data and simulation~\cite{CMS:2016lmd}.
The jet energy resolution amounts typically to 15--20\% at 30\GeV, 10\% at 100\GeV, and 5\% at 1\TeV~\cite{CMS:2016lmd}.
Additional selection criteria are applied to each jet to remove jets potentially dominated by anomalous contributions from various subdetector components or reconstruction failures~\cite{CMS-PAS-JME-16-003}.

The missing transverse momentum vector \ptvecmiss is computed as the negative vector sum of the transverse momenta of all the PF candidates in an event, and its magnitude is denoted as \ptmiss~\cite{CMS:2019ctu}.
The \ptvecmiss is modified to account for corrections to the energy scale of the reconstructed jets in the event.

During the 2016-2017 data taking, a gradual shift in the timing of the inputs of the ECAL L1 trigger in the region at $\abs{\eta} > 2.0$ caused a specific trigger inefficiency~\cite{CMS:EGM-18-002}, called ``prefiring'' hereafter.
For events containing a jet with $\pt \gtrsim 100\GeV$ in the region $2.5 < \abs{\eta} < 3.0$, the efficiency loss is $\approx$10--20\%, depending on \pt, $\eta$, and data-taking period.

\section{Data analysis}
\label{sec:Measurement}

\subsection{Event selection criteria}
\label{subsec:Select}

{\tolerance=800 Each event is required to have at least one offline-reconstructed PV with $z$ coordinate satisfying the criterion $\abs{z_\mathrm{PV}} <24\cm$ and radial distance from the interaction point $\rho_{\mathrm{PV}}<2\cm$.\par}
Anomalous high-\ptmiss events can be due to a variety of reconstruction  failures, detector malfunctions, or noncollision backgrounds.
Such events are rejected by event filters that are designed to identify more than 85--90\% of the spurious high-\ptmiss events with a mistagging rate less than 0.1\%~\cite{CMS:2019ctu}.
Only events that have been accepted by at least one single-jet trigger path (described in Section~\ref{subsec:Trigger}) are included in the measurement.
For the rejection of poorly reconstructed jets and jets originating from detector noise, additional quality criteria are applied to them based on their constituents~\cite{CMS-PAS-JME-16-003}. 

The measurement is based on an inclusive jet sample that contains only jets reconstructed within the rapidity range $\yabs < 2.5$ and with transverse momenta $\pt>50\GeV$.
The \pTnbrmin threshold and azimuthal separation interval for neighbouring jets as defined in Eq.~(\ref{eq:observable}), are set to 100\GeV and $2\pi/3<\Delta\phi<7\pi/8$, respectively.
This choice was motivated by statistical optimization, extending the phase space as much as possible, and guaranteeing that the NLO predictions remain valid and soft effects are negligible.

\subsection{Triggers}
\label{subsec:Trigger}

The analysis is based on single-jet triggers that require at least one jet with \pt above a given threshold $(\pt^{\text{thresh}})$ to be present in the event.
Table~\ref{table:triggers} shows the different HLT thresholds along with the effective integrated luminosities recorded by the triggers for 2016, 2017 and 2018.
All triggers in Table~\ref{table:triggers} were prescaled, apart from trigger 450
for 2016, and trigger 500 for 2017 and 2018.
The efficiency for each trigger is estimated as a function of leading-jet \pt using a lower-threshold trigger, except for the lowest-threshold trigger efficiency.
The latter is estimated using a tag-and-probe method applied to dijet topologies, which counts the jets reconstructed with the offline PF algorithm that can be matched to HLT jets, considering only the jets leading and subleading in \pt.
The data consist of events selected with a combination of triggers in mutually exclusive leading-jet \pt intervals.
The usage of a specific trigger is enabled only in phase-space regions where its efficiency is larger than 99.5\% and disabled in phase-space regions where the efficiency of a higher-threshold trigger is larger than 99.5\%.
The jet-counting variables are combined event-by-event by applying weights to account for the trigger prescales of each data sample.
 
\begin{table*}[htbp]
	\centering
	\topcaption{The different HLT \pt thresholds used in the measurement and the corresponding integrated luminosities for each data-taking year.}
	\begin{tabular}{cccccccccccc}
	$\pt^{\text{thresh}}$ ($\GeVns{}$) &   \hspace{0.01cm}       & 40     & 60    & 80   & 140  & 200  & 260 & 320  & 400  & 450   & 500  \\
	\hline
	                        & 2016                            & 0.0497 & 0.328 & 1.00 & 10.1 & 85.8 & 518 & 1526 & 4590 & 33500 & \NA \\
  \lumi (\pbinvns)          & 2017                            & 0.182  & 0.505 & 2.53 & 26.6 & 189  & 469 & 1230 & 7690 & 9660 & 41500 \\
	                        & 2018                            & 0.0151 & 0.419 & 2.17 & 47.1 & 202  & 466 & 1240 & 3720 & 7390 & 59800 \\
	\end{tabular}
	\label{table:triggers}
\end{table*} 

\subsection{Unfolding}
\label{subsec:Unfold}

{\tolerance=800 To compare the experimental data with theoretical predictions, the measured distributions must be corrected for detector effects, such as finite \pt resolution and limited detector acceptance.
The detector effects are parameterized through a response matrix built from simulated event samples using \PYTHIA~8.240~\cite{bib:Pythia8} with tunes CUETP8M1~\cite{bib:CUETM1} and CP5~\cite{bib:CP5}, where reconstructed-level jets are matched to generator-level jets as explained next.
First, the generated jets in each event are ordered by decreasing \pt.
Then, each generated jet is matched to the reconstructed jet with the highest \pt, within a cone of radius $\Delta R= \sqrt{\smash[b]{(\Delta\eta)^2+(\Delta\phi)^2}} = 0.35$ (where $\Delta\eta$ and $\Delta\phi$ are the angular differences in pseudorapidity and azimuthal angle between the generated and reconstructed jets).
The probability matrix $\mathbold{A}$ corresponds to the row-by-row normalized response matrix.
Each element $A_{ij}$ represents the probability of a jet produced in (generator-level) bin $j$ to be observed in (reconstructed-level) bin $i$.
The detector effects are corrected through an unfolding procedure that accounts for bin migrations, background (fake jets, \ie, reconstructed-level jets that could not be matched to generator-level jets), and inefficiencies (missed jets, \ie, generator-level jets that could not be matched to reconstructed-level jets), and corrects the measurement from the detector level to the level of stable particles (except neutrinos) with mean decay-lengths larger than $c\tau=10\unit{mm}$ (where $\tau$ denotes the mean proper lifetime of the particle).\par}

{\tolerance=1200
The unfolding procedure is implemented using the \textsc{TUnfold} package~\cite{bib:TUnfold}.
The determination of the particle-level distribution ($\mathbold{x}$) is performed with the matrix pseudoinverse method~\cite{bib:DataAnalysisHEP} using a detector-level distribution ($\mathbold{y}$) with twice the number of bins of the particle-level distribution.
The latter are defined in Table~\ref{table:Systematics} (first column) and are chosen to ensure that the bin sizes remain at least twice as large as the jet \pt resolution. The unfolding solution arises from the minimization of the quantity
\begin{equation}
  \label{eq:unfold}
  \chisq = \mathbold{\left(Ax + b - y\right)^T (V^{-1}) \left(Ax + b - y\right)},
\end{equation}
where $\mathbold{b}$ is the background obtained from simulated events, and $\mathbold{V}$ is the covariance matrix (corrected for the background) of the detector-level data including their statistical uncertainties and correlations.
Instead of unfolding separately the numerator and denominator of Eq.~(\ref{eq:observable}), a multidimensional unfolding of the more general, equivalent quantity \Nptn is performed, which rigorously accounts for the numerator-denominator statistical correlations following Eq.~(\ref{eq:observableEquiv}).
\par}

\begin{figure*}[hbtp]
  \centering
  \includegraphics[width=0.98\textwidth]{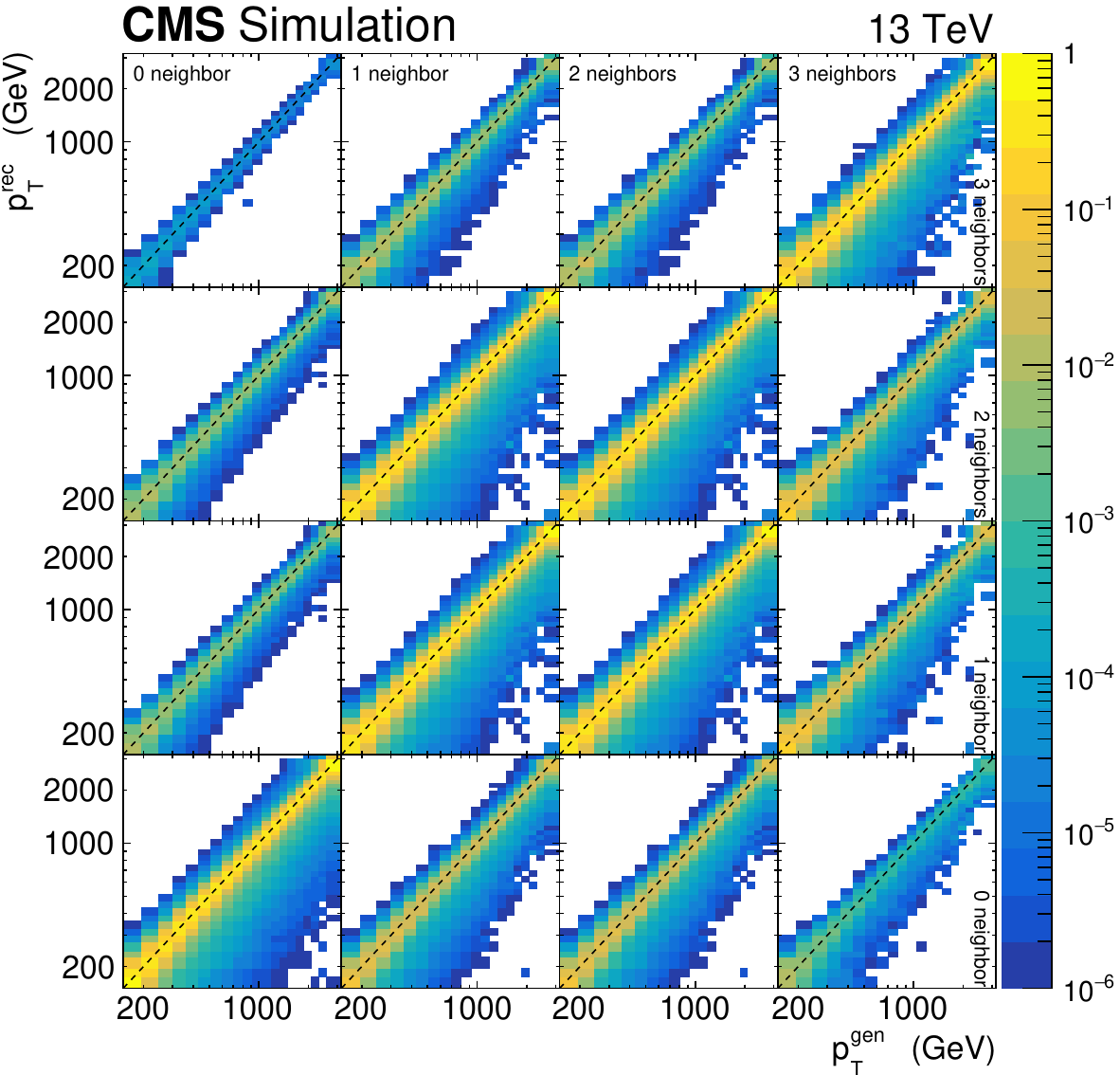}
  \caption{Probability matrix for the \Nptn distribution built using \PYTHIAEIGHT simulated events. The
    horizontal axis corresponds to the generator-level jet \pt, and the vertical axis to the reconstructed-level jet \pt. The
    $4\times 4$ structure of the matrix corresponds to the bins of neighbouring jets $n$ (labelled in the uppermost row and
    rightmost column), and indicates migrations among those bins. The horizontal and vertical axes of each cell
    correspond to the \pt of the jets, and each cell indicates the migrations among the jet \pt bins. 
    The range of colours covers from $10^{-6}$ to 1, and indicates the probability of migrations from a given
    (generator) particle-level bin to the corresponding (reconstructed) detector-level bin.
  \label{fig:ProbMatrix}}
\end{figure*}

Figure~\ref{fig:ProbMatrix} shows the probability matrix for the \Nptn quantity.
The number of events with $n\geq4$ is small, and $n=3$ is the maximum number of neighbouring jets shown here. 
The condition number (defined as the absolute value of the ratio between the largest and smallest matrix eigenvalues) for the matrix $\mathbold{A}$ is ${\approx}5.5$, which means that the unfolding problem is well-conditioned, and therefore no additional regularization is required.

\subsection{Experimental uncertainties}
\label{subsec:Uncertainties}

\begin{figure*}[hbt!]
  \centering
	\includegraphics[width=0.98\textwidth]{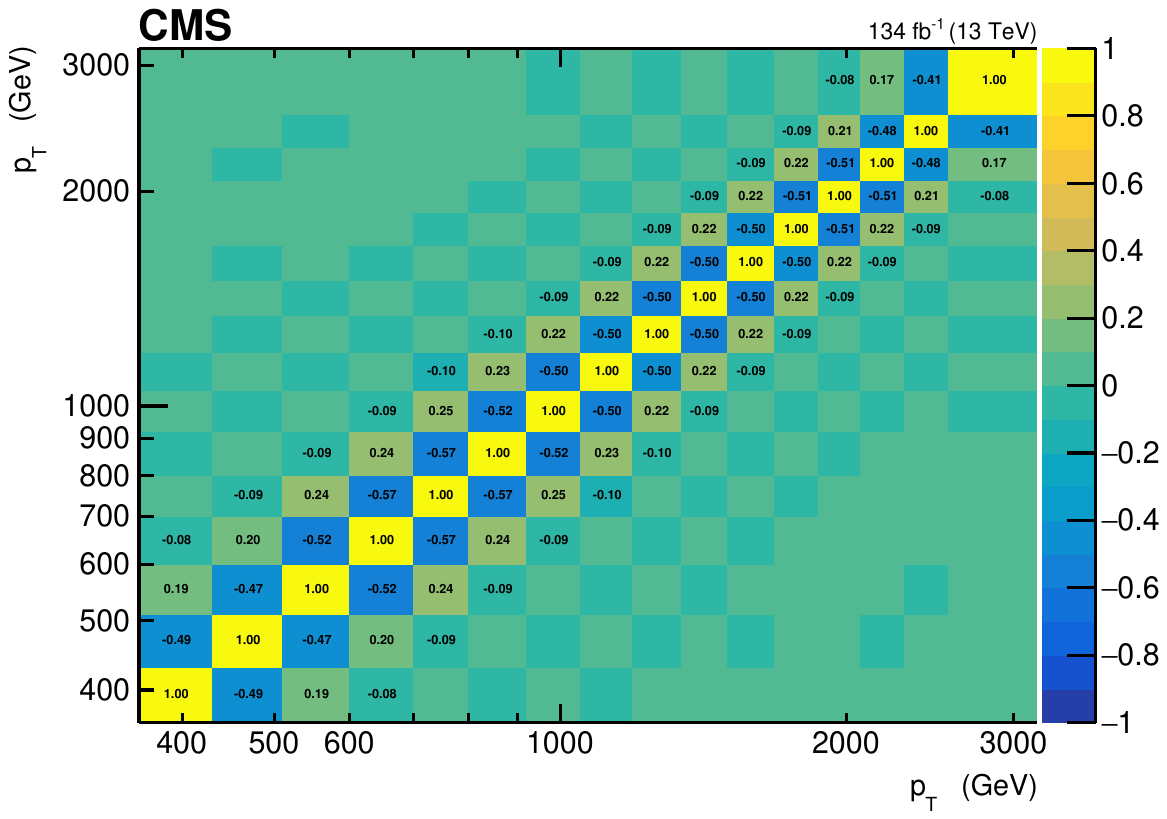}
	\caption{Bin-to-bin correlation matrix for the \rdphi distribution at the particle level, where the value 1~$(-1)$ corresponds
		     to fully (anti)correlated bins. For illustration purposes, only bins with (anti)correlations larger (smaller) than 
		     0.05~$(-0.05)$ are shown also as text.}
	\label{fig:Correlation}
\end{figure*}

{\tolerance=800 The experimental uncertainties contain statistical and systematic sources that propagate to the measured distributions.
The statistical uncertainties are obtained from the covariance matrix, extracted at the particle level from the \Nptn
distribution along with the \rdphi observable, as described in Section~\ref{subsec:Unfold}.
The bin-to-bin correlation matrix at the particle level is shown in Fig.~\ref{fig:Correlation}, where the value 1~$(-1)$
corresponds to fully (anti)correlated bins. The diagonal elements of the correlation matrix are by construction always unity, and
the off-diagonal elements represent the bin-to-bin (anti)correlations, where the highest (lowest) value is 0.49~($-0.57$).
The statistical uncertainties in the \rdphi measurement remain below 1\% up to ${\approx}1.5\TeV$ increasing to about 10\% at ${\approx}3\TeV$.\par}

{\tolerance=800 The calibration of the reconstructed jet energy is performed through a series of successive stages implemented in the
JES correction procedure~\cite{CMS:2016lmd}, as described in Section~\ref{sec:Reconstruction}. 
The JES uncertainty is composed of 27 individual uncorrelated contributions, which are investigated one-by-one considering a
$\pm1$ standard-deviation variation from their nominal value. Each variation is applied at the detector level
and propagated to the particle-level measurement by repeating the unfolding procedure. Finally, the total JES 
uncertainty is computed as the quadratic sum of individual JES uncertainty sources, and remains below 1\%
up to ${\approx}1.5\TeV$ increasing to about 5\% at ${\approx}3\TeV$. Additional variations of the trigger prefiring corrections
described in Section~\ref{sec:Reconstruction} are applied in the same manner. The uncertainties from prefiring
corrections in the \rdphi measurement are smaller than 0.13\%.\par}

{\tolerance=1800 In simulated samples, a detailed modelling of the CMS detector is included based on the \GEANTfour toolkit~\cite{bib:Geant4}.
The JER obtained in the detector simulation is generally better than that in the actual detector. Therefore, 
the energy of reconstructed jets in simulation is smeared out, so that the simulated JER matches the one
measured in experimental data. The JER uncertainty is estimated by varying the smearing factors within $\pm1$ standard deviations from their
nominal values, and propagated to the particle-level measurement by repeating the unfolding procedure. The JER uncertainty
in the \rdphi measurement is below 0.8\%.\par}

The probability distribution for the number of $\Pp\Pp$ interactions per bunch crossing is represented by pileup (PU) profiles.
To account for differences between the measured and simulated PU profiles, the simulated events are reweighted using
the PU distribution of the experimental data as a reference. An additional systematic uncertainty, which remains below 0.03\%, is evaluated
by varying the PU profile correction in the simulation. The model dependence introduced by unfolding is estimated from the
difference in the \rdphi distribution unfolded with response matrices obtained from \PYTHIAEIGHT and \MGvATNLO~2.6~\cite{Alwall:2014hca, Alwall:2007fs} interfaced with \PYTHIAEIGHT. Additionally, the model dependence for inefficiencies (missed jets) and backgrounds (misreconstructed jets) is studied by varying separately their rates within an estimate of 5\%, which largely covers the model dependence of migrations in and out of the phase space. The total model dependence uncertainties ($\text{MC}_{\text{model}}$) are negligible (${<}0.01\%$) compared with other uncertainties for the bulk of the spectrum. The trigger efficiency uncertainties are also negligible in the \rdphi measurement. The \rdphi observable values along with all the experimental uncertainties are shown in Table~\ref{table:Systematics}.

\begin{table*}[htpb]
	\centering
	\topcaption{Values of the \rdphi observable in different \pt intervals, and associated experimental uncertainties.}
	\begin{tabular}{cccccccc}
		\pt ($\GeVns$) & \rdphi & Stat. (\%)  & JES (\%) & Prefiring (\%) & JER (\%) & PU (\%) & $\text{MC}_{\text{model}}$ (\%)\\
		\hline
        360--430             & 0.25             & 0.26                      & 0.74             & 0.06                   & 0.08             & 0.01            &  \NA \\          
        430--510             & 0.26             & 0.23                      & 0.69             & 0.09                   & 0.06             & 0.02            &  \NA \\           
        510--600             & 0.27             & 0.19                      & 0.67             & 0.12                   & 0.05             & 0.02            &  \NA \\     
        600--700             & 0.27             & 0.18                      & 0.66             & 0.13                   & 0.04             & 0.02            &  0.01 \\     
        700--800             & 0.28             & 0.18                      & 0.65             & 0.12                   & 0.04             & 0.02            &  0.01 \\     
        800--920             & 0.28             & 0.20                      & 0.65             & 0.10                   & 0.04             & 0.02            &  0.01 \\    
        920--1050            & 0.27             & 0.27                      & 0.66             & 0.07                   & 0.05             & 0.02            &  \NA \\     
        1050--1190           & 0.27             & 0.39                      & 0.67             & 0.05                   & 0.06             & 0.02            &  \NA \\     
        1190--1340           & 0.27             & 0.58                      & 0.70             & 0.03                   & 0.07             & 0.02            &  \NA \\    
        1340--1500           & 0.26             & 0.90                      & 0.75             & 0.02                   & 0.08             &	0.02            &  \NA \\    
        1500--1680           & 0.26             & 1.34                      & 0.84             & 0.01                   & 0.11             &	0.02            &  \NA \\   
        1680--1870           & 0.25             & 2.16                      & 0.98             & \NA                      & 0.14             &	0.02            &  \NA \\     
        1870--2070           & 0.24             & 3.42                      & 1.21             & \NA                      & 0.19             & 0.02            &  \NA \\     
        2070--2300           & 0.21             & 5.66                      & 1.62             & \NA                      & 0.27             & 0.02            &  \NA \\    
        2300--2560           & 0.22             & 8.23                      & 2.36             & \NA                      & 0.39             & 0.02            &  \NA \\   
        2560--3170           & 0.21             & 10.49                     & 5.00             & \NA                      & 0.77             & 0.01            &  \NA \\    
	\end{tabular}
	\label{table:Systematics}
\end{table*}

\section{Theoretical predictions}
\label{sec:TheoreticalPredictions}

\subsection{Monte Carlo event generators predictions}
\label{subsec:MonteCarlo}

{\tolerance=1800
Experimental data are compared with predictions from \HERWIGpp~2.7.1~\cite{bib:Herwig}, \PYTHIA~8.240~\cite{bib:Pythia8}, and
\POWHEG~2.0~\cite{bib:Powheg2} Monte Carlo (MC) event generators obtained using the \textsc{Rivet}
toolkit~\cite{bib:Rivet}. The \HERWIGpp event generator computes the matrix elements (MEs) at
LO accuracy for $2 \to 2$ QCD scattering processes. The parton shower (PS) is simulated through successive angular-ordered emissions, 
and the cluster fragmentation model~\cite{bib:Cluster} is used for the hadronization. The underlying event (UE) activity is obtained 
from the simulation of multiparton interactions (MPIs) tuned to experimental data. The set of \HERWIGpp parameters used in this analysis 
is that of the UE-EE-5-CTEQ6L1 tune~\cite{bib:EE5C} based on the CTEQ6.1M LO PDF
set~\cite{bib:CTEQLO}. Similarly to \HERWIGpp, in \PYTHIAEIGHT the MEs are calculated at LO accuracy
for $2 \to 2$ QCD scattering processes. The PS is simulated through successive \pt-ordered emissions and the
hadronization mechanism employs the Lund string model~\cite{bib:Lund}. Two different sets of parameters are used
for \PYTHIAEIGHT, the CUETP8M1 tune~\cite{bib:CUETM1} based on the NNPDF2.3 LO PDF
set~\cite{bib:NNPDF2.3-1,bib:NNPDF2.3-2} and the CUETP8M2 tune~\cite{bib:CUETM2} based on the NNPDF3.0
LO PDF set~\cite{bib:NNPDF30}. The \POWHEG~\cite{bib:Powheg0, bib:Powheg1} generator, based on the \textsc{Powheg
  box}~\cite{bib:Powheg2}, generates $2 \to 2$ matrix elements at NLO accuracy, as well as
$2 \to 3$ matrix elements at LO accuracy and uses the NNPDF3.0 NLO PDF
set~\cite{bib:NNPDF30}. To simulate the PS, hadronization, and MPI processes, \POWHEG is interfaced
either with \PYTHIAEIGHT or with \HERWIGpp.
\par}

\begin{figure*}[hbt!]
  \centering
  \includegraphics[width=0.98\textwidth]{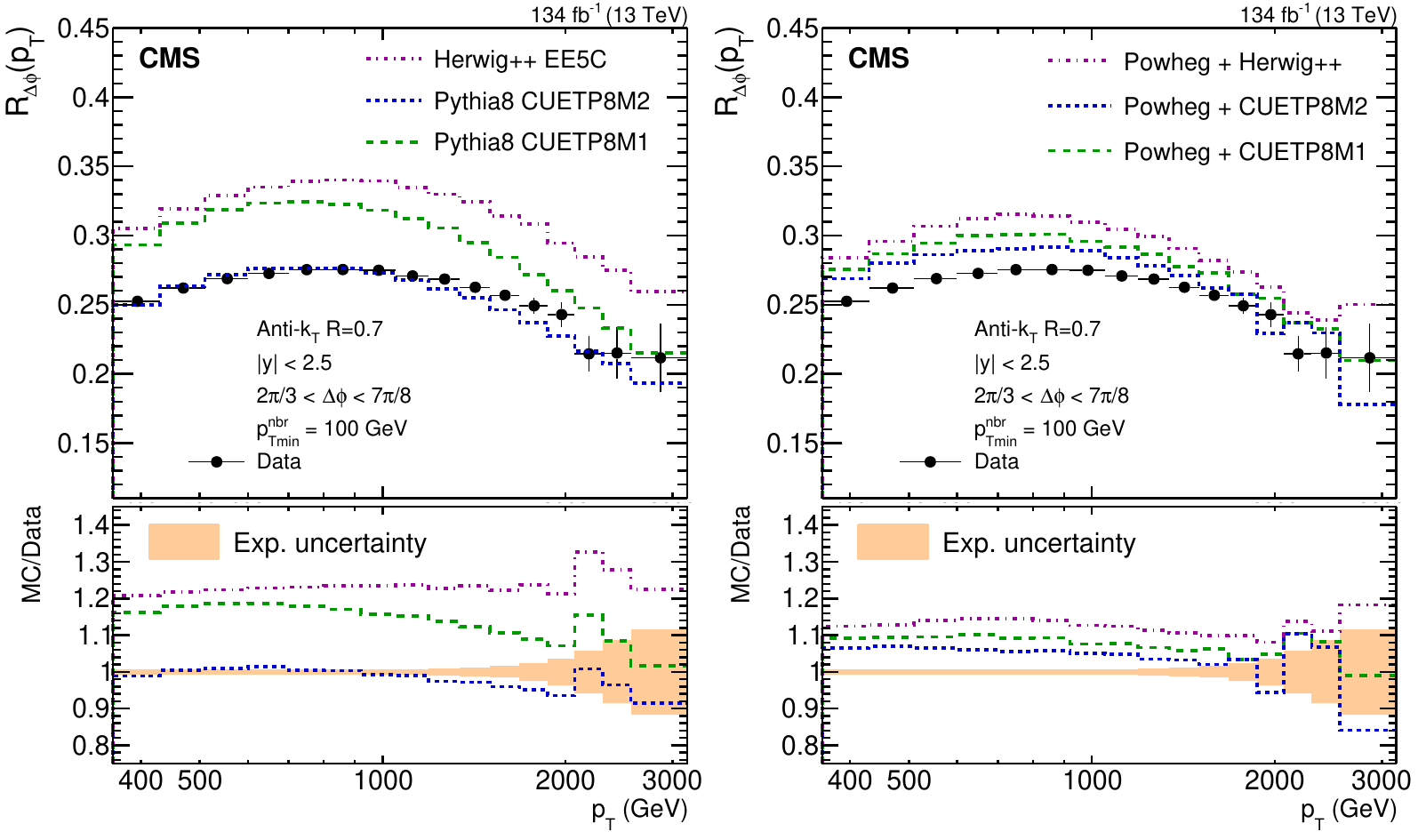}
  \caption{The \rdphi observable as a function of \pt, compared with MC generator predictions at
    LO (left) and at NLO (right) accuracy. The LO predictions are obtained with \PYTHIAEIGHT
    tunes CUETP8M1 and CUETP8M2, and \HERWIGpp tune
    UE-EE-5-CTEQ6L1 MC event generators. The NLO predictions are obtained with \POWHEG
    interfaced with each of the aforementioned MC event generators. The experimental data are represented with black dots and the
    MC predictions with coloured lines. The lower panel of each plot shows the ratio between MC predictions and experimental data.
    The total experimental uncertainties are indicated by the vertical error bars (upper panels) and coloured band
    (lower panels) correspondingly.}
  \label{fig:MonteCarlo}
\end{figure*}

{\tolerance=800 Figure~\ref{fig:MonteCarlo}~(left) shows the particle-level data for the \rdphi observable compared with the predictions of
\PYTHIAEIGHT tunes CUETP8M1 and CUETP8M2, and \HERWIGpp tune UE-EE-5-CTEQ6L1 LO MC event generators. 
The lower panels show the corresponding ratios between the MC predictions and the measured data.
Figure~\ref{fig:MonteCarlo}~(right) illustrates the particle-level \rdphi observable compared with \POWHEG interfaced with \PYTHIAEIGHT tunes CUETP8M1 and CUETP8M2, and \HERWIGpp tune UE-EE-5-CTEQ6L1 results. The corresponding lower panels show the ratios between the \POWHEG predictions and the measurement. 
The coloured band on both lower panels represents the total experimental uncertainties.\par}

From 360 up to around 800\GeV, the \rdphi distribution rises as the phase space for the production of a
third jet increases. Then, the distribution reaches a plateau up to around 1200\GeV, followed by a decrease due to the
running of \alpS , and the reduced amount of gluon scatterings.

The predictions from LO \HERWIGpp and LO \PYTHIAEIGHT tune CUETP8M1 overestimate the measurement by ${\approx}20\%$ and ${\approx}12$--18\%, respectively.
On the other hand, the predictions from the (LO) \PYTHIAEIGHT tune CUETP8M2 give a good description of the data. Besides the PDF set, the main differences
between the parameters of the two \PYTHIAEIGHT tunes are the value of \alpS used for the initial-state shower $\alpS^{\text{ISR}}$, the MPI infrared regularization
scale $p_{\text{T0}}^{\text{ref}}$, and the amount of colour reconnection. Among the NLO MC predictions based on \POWHEG, the \POWHEG interfaced with \PYTHIAEIGHT tune
CUETP8M2 gives the best description, being ${\approx}5$--6\% away from the measurement. Finally, \POWHEG interfaced with \HERWIGpp tune UE-EE-5-CTEQ6L1 or with
\PYTHIAEIGHT tune CUETP8M1 overestimate the \rdphi measurement by ${\approx} 12\%$ and ${\approx}10\%$, respectively.

\subsection{Fixed-order predictions}
\label{subsec:FixedOrder}

{\tolerance=1200
Fixed-order theoretical predictions for the \rdphi observable are obtained up to NLO accuracy in pQCD with the \NLOJETPP
program~\cite{bib:NLOJET1,bib:NLOJET2} within the \fastNLO framework~\cite{bib:fastNLO1,bib:fastNLO2}. The predictions
are extracted for several PDF sets available via the \textsc{lhapdf} library~\cite{bib:LHAPDF}, using their
default value for the strong coupling constant \alpsmz, and alternative values, as shown in Table~\ref{table:PDFs}. The
central reference values $\mu_0$ for the renormalization (\mur) and factorization (\muf) scales are defined as:
\begin{equation}
  \label{eq:Scales}
  \mur=\muf=\hatHTtwo,
\end{equation}
where \hatHT is the scalar sum of the transverse momenta of all partons in the event. This choice follows
recommendations detailed in Ref.~\cite{Currie:2018xkj}, which favour \hatHT over $p_{\mathrm{T},\text{jet}}$ and
discourage the use of $p_{\mathrm{T},\text{max}}=p_\mathrm{T,1}$ as the central scale choice for inclusive jet cross sections. 
For 3-jet ratio observables such as \rdphi, Refs.~\cite{Czakon:2021mjy,Alvarez:2023fhi} conclude that \hatHTtwo is slightly preferred for comparisons with theoretical predictions at
next-to-NLO accuracy.
The uncertainties related to missing higher-order terms of the perturbative series are estimated using the conventional recipe~\cite{Cacciari:2003fi,Catani:2003zt,Banfi:2010xy}, \ie,
by varying \mur and \muf around the reference scale $\mu_{0}$ within six combinations: $(\mur/\mu_{0},\muf/\mu_{0})=(1/2,1/2)$, $(1/2,1)$, $(1,1/2)$, $(1,2)$, $(2,1)$, $(2,2)$.
An envelope is constructed from the various combinations, where the edges define the scale uncertainties.
\par}

\begin{table}[htpb]
  \centering 
  \topcaption{Default and range of \alpsmz values used in the different NLO PDF sets.}
  \begin{tabular}{lcc}
    PDF set & Default \alpsmz  & Alternative \alpsmz \\
    \hline
    {ABMP16}~\cite{bib:ABMP16}   & 0.1191   & 0.114--0.123      \\
    {CT18}~\cite{bib:CT18}            & 0.1180   & 0.110--0.124      \\
    {MSHT20}~\cite{bib:MSHT20}        & 0.1200   & 0.108--0.130      \\
    {NNPDF3.1}~\cite{bib:NNPDF}      & 0.1180   & 0.106--0.130      \\
  \end{tabular}
  \label{table:PDFs}
\end{table}

\begin{figure*}[hbt!]
  \centering
  \includegraphics[width=0.98\textwidth]{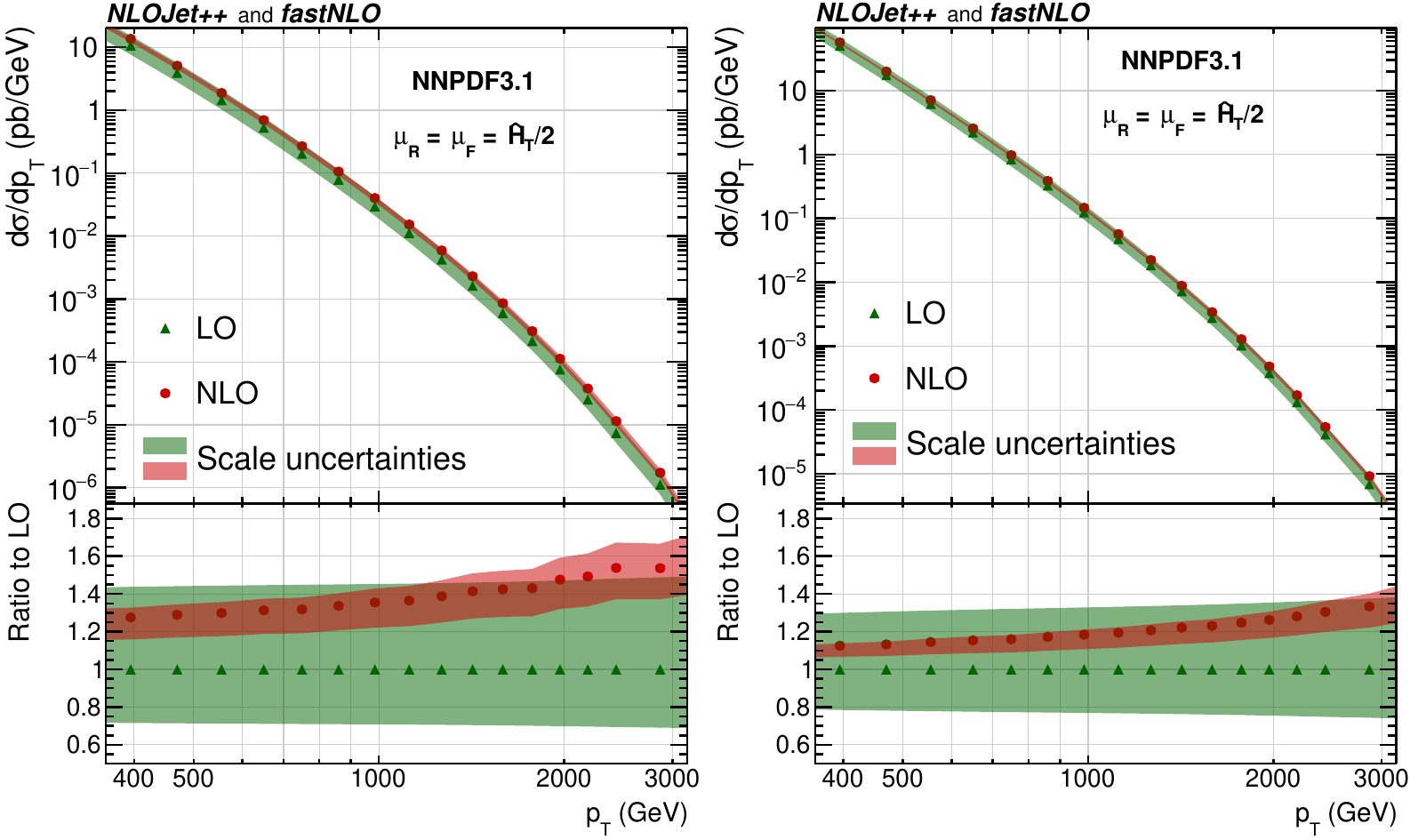}
  \caption{Theoretical predictions for the cross sections corresponding to the numerator (left) and denominator (right) of the \rdphi ratio, Eq.~(\ref{eq:observable}),
   obtained using the NNPDF3.1 NLO PDF set. The coloured bands represent the LO and NLO scale uncertainties derived with a six-point
   variation of \mur and \muf from the central reference value. The lower panels show the ratios to the respective LO predictions.}
  \label{fig:TheoryNNPDF31ScaleHT2}
\end{figure*}

Theoretical calculations are performed separately for the cross sections corresponding to the jet counts in the numerator and denominator of the \rdphi ratio defined in Eq.~(\ref{eq:observable}). The predictions using NNPDF3.1 for the numerator (left) and denominator (right) differential cross sections ($\rd\sigma/\rd\pt$) at LO and NLO accuracy are shown in Fig.~\ref{fig:TheoryNNPDF31ScaleHT2}, along with the scale uncertainties (coloured bands).
The lower panels in this figure display the ratios to the respective LO predictions, where the so-called NLO pQCD $K$ factors (NLO/LO) are about 1.30--1.55 for the numerator
and 1.20--1.35 for the denominator (1.08--1.15 for their ratio). The LO and NLO scale uncertainty bands overlap over the whole phase space. The NLO scale uncertainties are in
the range 9--17\% for the numerator and 5--10\% for the denominator.

\begin{figure*}[hbtp]
  \centering
  \includegraphics[width=0.98\textwidth]{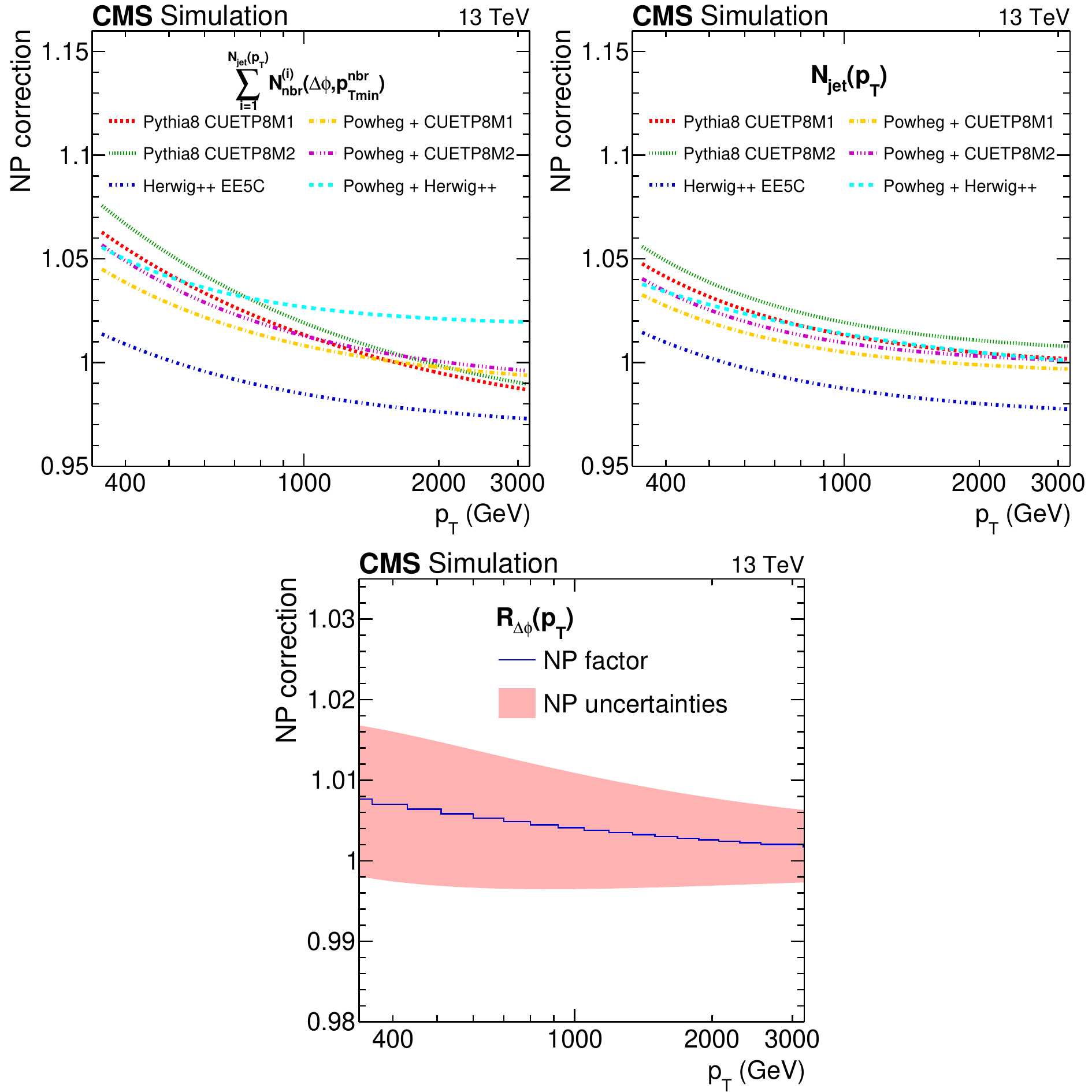}
  \caption{Nonperturbative correction factors for the numerator (upper left) and denominator (upper right)
    of the \rdphi ratio, Eq.~(\ref{eq:observable}), using \PYTHIAEIGHT with tunes CUETP8M1 and 
    CUETP8M2,  \HERWIGpp with tune UE-EE-5-CTEQ6L1, and POWHEG interfaced with each of them. 
    The lower plot shows the NP correction factors (blue line) for \rdphi and their uncertainties.}
  \label{fig:NPcorrections}
\end{figure*}

To compare fixed-order predictions at parton level with unfolded data, the former must be corrected for NP effects due to MPI and hadronization (HAD).
Based on PS generators, the NP correction factors are evaluated from the ratio of the nominal over the generated cross sections when MPI and HAD effects are switched off:
\begin{equation}
  \label{eq:NP}
  C^{\text{NP}}=\frac{\sigma^{\text{PS+MPI+HAD}}}{\sigma^{\text{PS}}}.
\end{equation}
The model dependence of $C^{\text{NP}}$ is investigated using different MC event generators, namely \PYTHIAEIGHT with tunes CUETP8M1 and CUETP8M2, \HERWIGpp with
tune UE-EE-5-CTEQ6L1, and \POWHEG interfaced with each one of them. A simple polynomial function $a + b \pt^c$ (where $a$, $b$, and $c$ are free parameters) is used to
parameterize the dependence of $C^{\text{NP}}$ on jet \pt for each MC event generator, to avoid statistical fluctuations in less populated regions of phase space. An envelope
is constructed from the different MC predictions, where the central values are identified as the NP correction factors $C^{\text{NP}}$ and the edges define the corresponding uncertainties.
Figure~\ref{fig:NPcorrections} shows the NP correction factors obtained for the numerator (upper left) and denominator (upper right) of the \rdphi observable. The lower panels show the final
NP correction factors $C^{\text{NP}}$ (blue line) for \rdphi. The red band is constructed from the envelope of individual ratios and represents the relevant uncertainties, which are less
than 1\%.

\begin{figure}[hbt!]
  \centering
  \includegraphics[width=\cmsFigWidthA]{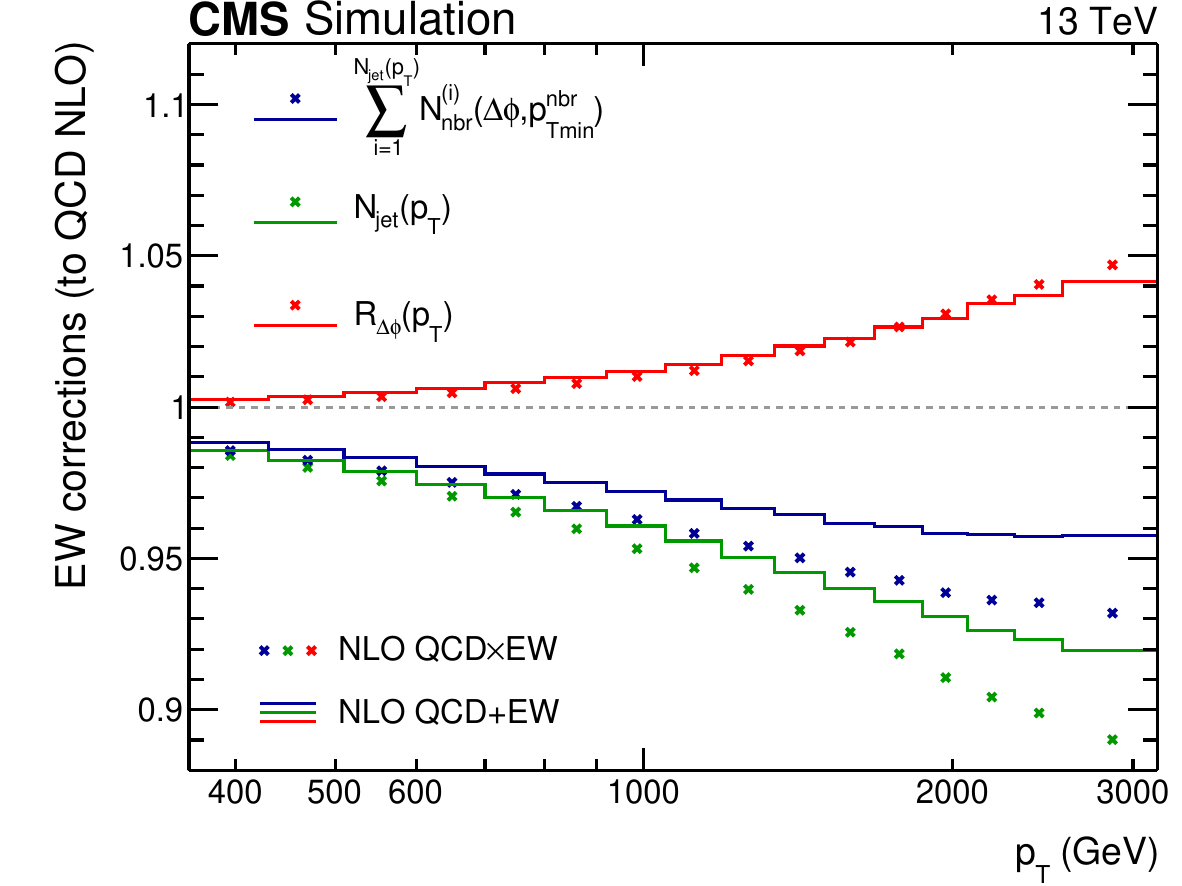}
  \caption{Electroweak corrections for the numerator (blue) and denominator (green) of Eq.~(\ref{eq:observable}), and 
  for the \rdphi ratio itself (red). The solid lines correspond to the additive combination of NLO EW corrections 
  to the QCD process (NLO QCD$\,+\,$EW), and the markers represent the multiplicative combination (NLO QCD$\,\times\,$EW).}
  \label{fig:EW}
\end{figure}

To further improve the accuracy, in particular at large jet \pt, the theoretical predictions are complemented with
electroweak (EW) corrections. The complete set of NLO corrections for three-jet production at the LHC is presented in Ref.~\cite{bib:EW3jet}. To obtain the EW corrections for the \rdphi
observable, the \SHERPA event generator~\cite{bib:Sherpa} is used, interfaced with \textsc{recola}~\cite{bib:Recola1, bib:Recola2}. Further details on the implementation of the above
interface as well as on the method used for the subtraction of NLO EW infrared divergences are reported in Refs.~\cite{bib:SherRec1} and~\cite{bib:SherRec2}, respectively. The pure NLO
EW corrections for $n$-jet production are defined as:
\begin{equation}
  \label{eq:EW1}
  \sigma_{\mathrm{nj}}^{\text{NLO\ EW}} = \sigma_{\mathrm{nj}}^{\text{LO}} + \sigma_{\mathrm{nj}}^{\Delta\text{NLO}_1},
\end{equation}
where $\sigma_{\mathrm{nj}}^{\text{LO}}$ is the pure LO pQCD cross section and $\Delta\text{NLO}_1$ accounts for the virtual and
real EW corrections. The additive and multiplicative combination of the above corrections to the cross sections are defined, respectively, as:
\begin{equation}
  \label{eq:EW2}
  \sigma_{\mathrm{nj}}^{\text{NLO\ QCD+EW}} = \sigma_{\mathrm{nj}}^{\text{LO}} + \sigma_{\mathrm{nj}}^{\Delta\text{NLO}_0} + \sigma_{\mathrm{nj}}^{\Delta\text{NLO}_1},
\end{equation}
\begin{equation}
  \label{eq:EW3}
  \sigma_{\mathrm{nj}}^{\text{NLO\ QCD}\times\text{EW}} = \sigma_{\mathrm{nj}}^{\text{LO}}\left(1+\frac{\sigma_{\mathrm{nj}}^{\Delta\text{NLO}_0}}{\sigma_{\mathrm{nj}}^{\text{LO}}}\right)
  \left(1+\frac{\sigma_{\mathrm{nj}}^{\Delta\text{NLO}_1}}{\sigma_{\mathrm{nj}}^{\text{LO}}}\right),
\end{equation}
where $\Delta\text{NLO}_0$ accounts for the virtual and real QCD corrections. Figure~\ref{fig:EW} shows the EW
corrections obtained for the numerator and denominator cross sections, and for the \rdphi ratio. The multiplicative
combination, Eq.~(\ref{eq:EW3}), is considered as the main result, whereas the additive combination, Eq.~(\ref{eq:EW2}) 
is used as an uncertainty estimate. The relative change of the central \alpsmz result (Section~\ref{sec:AlphaS}), 
when the additive combination is used as the main result, is smaller than 0.2\%. 
The EW corrections for \rdphi observable range from 0.2 to 5.0\% and their relevant uncertainties from 0.01 to 0.53\%.

\begin{figure*}[hbtp]
  \centering \includegraphics[width=0.9\textwidth]{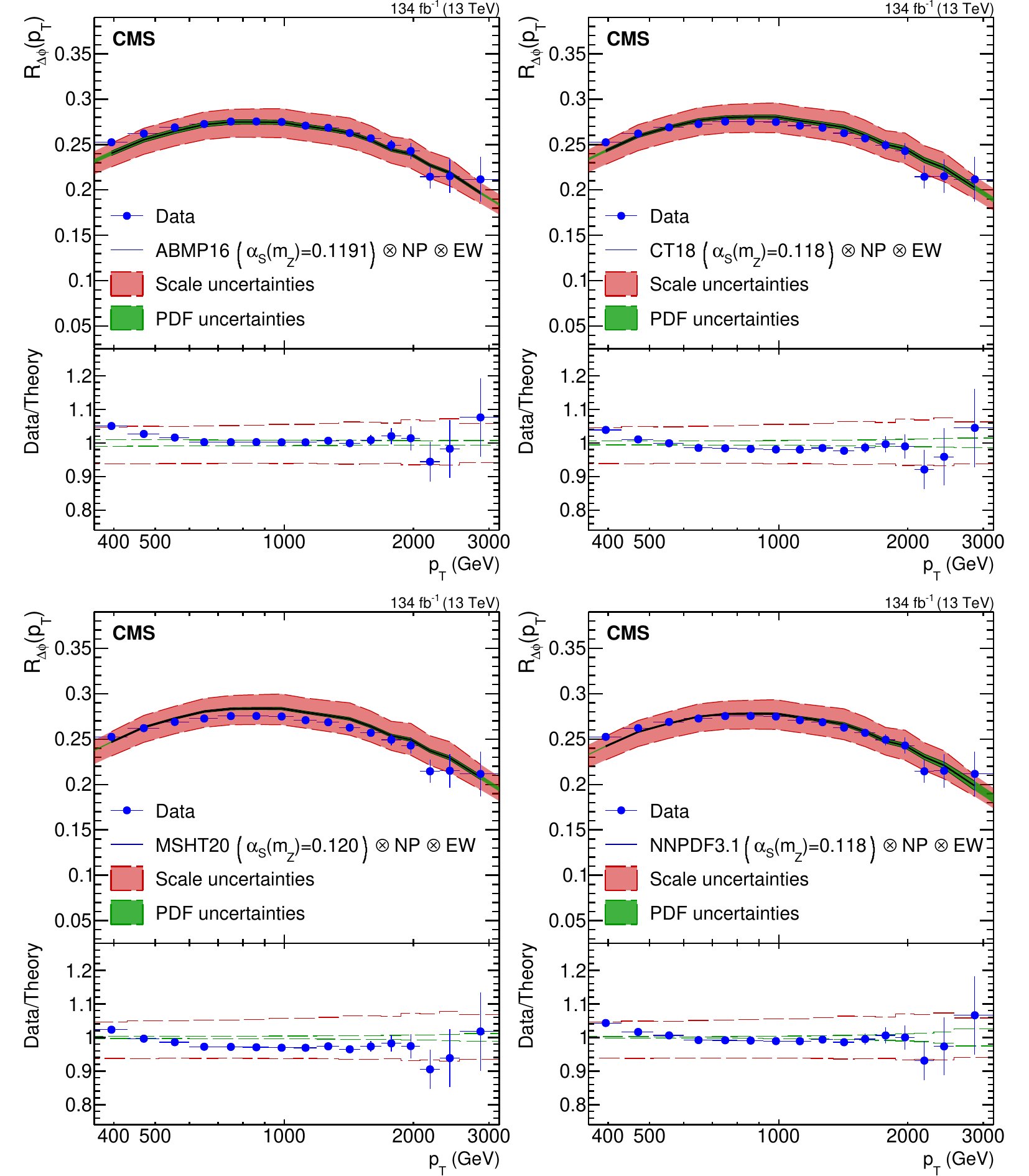}
  \caption{The \rdphi observable as a function of \pt, compared with fixed-order theoretical calculations at NLO accuracy using
    the ABMP16 (upper left), CT18 (upper right), MSHT20 (lower left), and NNPDF3.1 (lower right) NLO PDF sets. The experimental data
    are indicated with blue dots (with error bars representing the total experimental uncertainty), the theoretical 
    prediction for the default \alpsmz for each PDF set with black solid lines, the scale uncertainties with red bands, 
    and the PDF uncertainties with green bands. The lower panel of each plot
    shows the ratio between experimental data and theoretical predictions.}
  \label{fig:TheoryDataScaleHT2}
\end{figure*}

Comparisons between the measurement and the theoretical predictions for the four different PDF sets are shown in
Fig.~\ref{fig:TheoryDataScaleHT2}. The PDF uncertainties in the \rdphi predictions are evaluated at 68\% confidence
level for each PDF set following either the Hessian~\cite{bib:Hessian} or the MC~\cite{bib:MCmethod} methods, and are about 1--2\% in all cases.
The scale uncertainties in \rdphi predictions are dominant, ranging from 2 to 8\%. In general, all predictions (based on the default \alpsmz for each PDF set)
are in agreement with the measurement within the experimental and theoretical uncertainties.

\section{\texorpdfstring{Determination of $\alpsmz$}{Determination of alphas(mZ)}}
\label{sec:AlphaS}

\begin{figure*}[hbtp]
  \centering
  \includegraphics[width=0.98\textwidth]{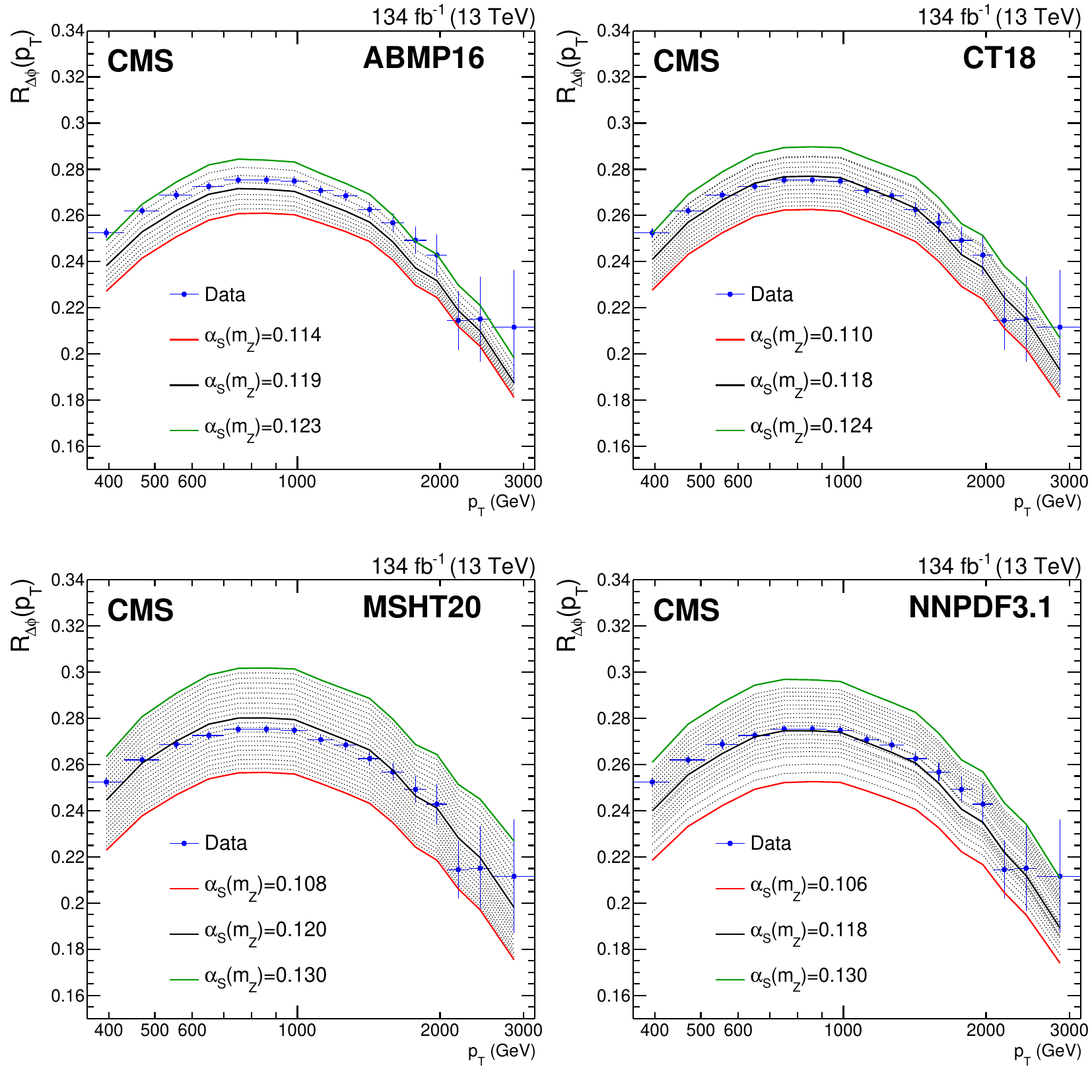}
  \caption{Sensitivity of the \rdphi ratio to the strong coupling constant \alpsmz. The data are indicated with blue dots
    with error bars representing the total experimental uncertainty. In each plot, the lines represent
    fixed-order NLO theoretical calculations obtained with ABMP16 (upper left), CT18 (upper right), MSHT20 (lower left)
    and NNPDF3.1 (lower right) NLO PDF sets. Solid green (red) lines indicate maximum (minimum) values, and dotted
    black lines intermediate values of \alpsmz for each PDF set.}
  \label{fig:SensitivityScaleHT2}
\end{figure*}

The sensitivity of the \rdphi ratio to the strong coupling constant is investigated by varying \alpsmz for each PDF set
within the ranges presented in Table~\ref{table:PDFs}. The \alpsmz value in the fixed-order matrix elements calculations is also adjusted accordingly.
Figure~\ref{fig:SensitivityScaleHT2} shows the results for each PDF set, where the solid green (red) curves
represent the maximum (minimum) \alpsmz values, and the dashed black curves correspond to intermediate \alpsmz values in $\Delta\alpsmz = \pm 0.001$ or $\Delta\alpsmz = \pm 0.002$ steps.
A large sensitivity of \rdphi to variations of the strong coupling constant is observed for all PDF sets, and hence \rdphi can be used for the determination of \alpsmz.

The value of \alpsmz is determined by minimising the goodness-of-fit (\chisq) between the experimental measurements and the theoretical
predictions. The \chisq is defined as:
\begin{equation}
  \label{eq:chisq}
  \chisq = \sum\limits_{ij}^N(D_{i}-T_{i})C_{ij}^{-1}(D_{j}-T_{j}),
\end{equation}
where $N$ is the number of measurements, $D_{i}$ are the experimental measurements, $T_{i}$ are the theoretical
predictions and $C_{ij}$ is the covariance matrix, which is composed of:
\ifthenelse{\boolean{cms@external}}
{
  \begin{multline}
    \label{eq:covar}
    C=C_{\text{stat}}+C_{\text{uncor}}+\left(\sum\limits_{\text{sources}}C_{\text{JES}}\right)
    +C_{\text{unfold}}\\+C_{\text{pref}}+C_{\text{NP}}+C_{\text{PDF}}+C_{\text{EW}},
  \end{multline}
}
{
\begin{equation}
  \label{eq:covar}
  C=C_{\text{stat}}+C_{\text{uncor}}+\left(\sum\limits_{\text{sources}}C_{\text{JES}}\right)
  +C_{\text{unfold}}+C_{\text{pref}}+C_{\text{NP}}+C_{\text{PDF}}+C_{\text{EW}},
\end{equation}
}
where $C_{\text{stat}}$ represents the statistical uncertainty, $C_{\text{uncor}}$ is the numerical precision of the
fixed-order predictions, which is assigned as uncorrelated uncertainty to each bin, $C_{\text{JES}}$ is the systematic
uncertainty for each JES uncertainty source,
$C_{\text{unfold}} = C_{\text{JER}} + C_{\text{PU}} + C_{\text{MC}_\text{model}}$ is the systematic uncertainty induced
through unfolding (representing the JER, pileup, and model uncertainties, respectively, described in
Section~\ref{subsec:Uncertainties}), $C_{\text{pref}}$ is the trigger prefiring uncertainty~\cite{CMS:EGM-18-002}, and $C_{\text{NP}}$,
$C_{\text{PDF}}$, and $C_{\text{EW}}$ are the NP, PDF, and EW uncertainties, respectively.

\begin{figure}[hbtp]
  \centering
  \includegraphics[width=\cmsFigWidthA]{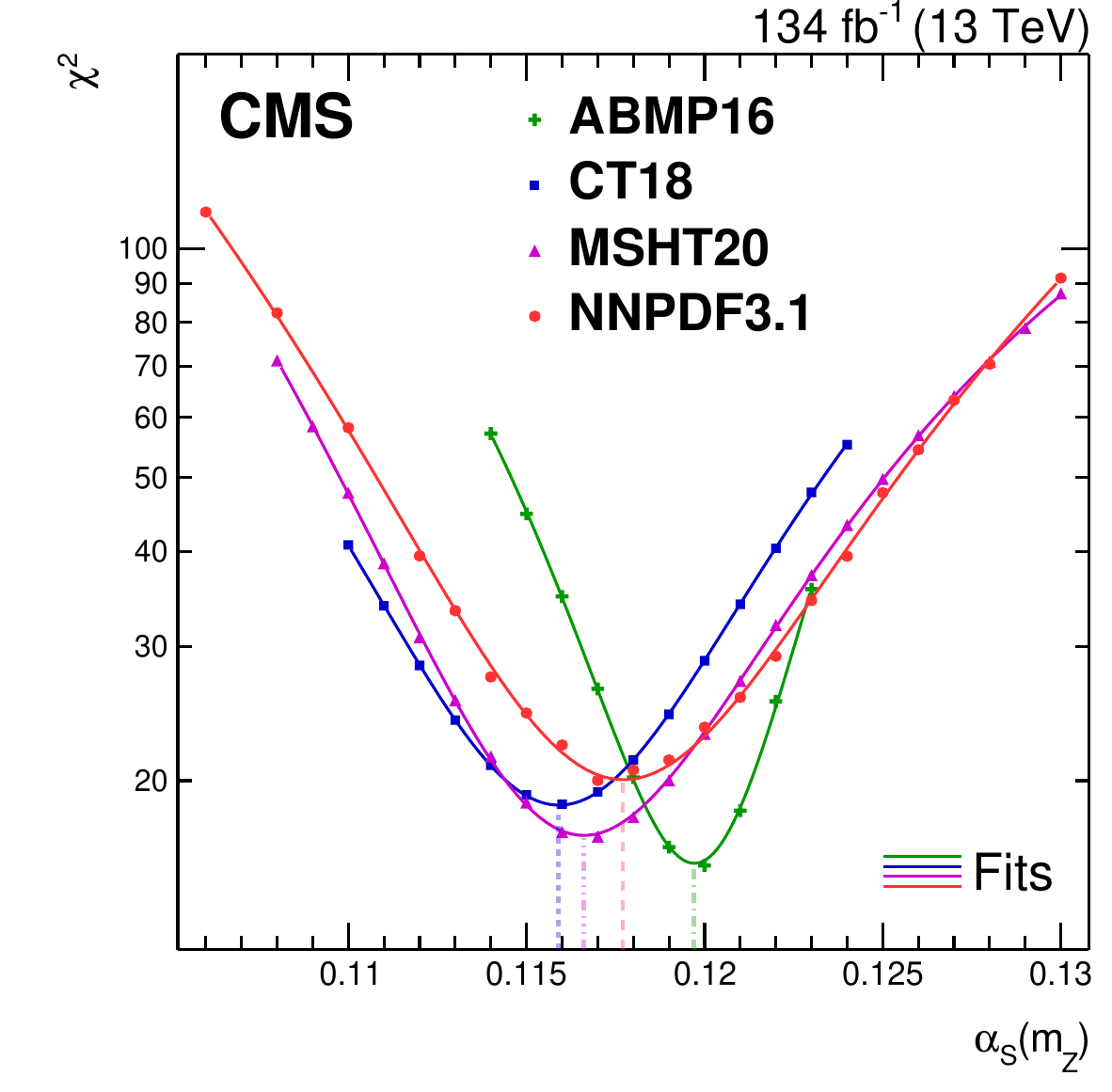}
  \caption{Minimization of the \chisq between experimental measurements and theoretical predictions for the \rdphi ratio, 
  with respect to \alpsmz for the ABMP16, CT18, MSHT20, and NNPDF3.1 NLO PDF sets. In this plot, only experimental uncertainties are
    included in the covariance matrix. The minimum value of \alpsmz found for each PDF set is indicated with a dashed
    line and corresponds to the central result. The experimental uncertainty is estimated from the \alpsmz values for which
    the \chisq is increased by one unit with respect to the minimum value.}
  \label{fig:AlphaSFit}
\end{figure}

The JES, unfolding, prefiring, NP, PDF, and EW uncertainties are considered as 100\% correlated among \pt bins, and are treated as multiplicative. Including only the experimental
(statistical, JES, unfolding, and prefiring) uncertainties in the covariance matrix composition, the central \alpsmz result is obtained by minimising the \chisq with respect to \alpsmz.
The associated experimental uncertainty is estimated from the \alpsmz values, for which the \chisq is increased by one unit with respect to the minimum
value. Figure~\ref{fig:AlphaSFit} illustrates the \chisq minimization curves for each PDF set, which result in the
\alpsmz values and their respective experimental uncertainties listed in Table~\ref{table:AlphasMz}.

\begin{figure}[hbtp]
  \centering
  \includegraphics[width=\cmsFigWidthA]{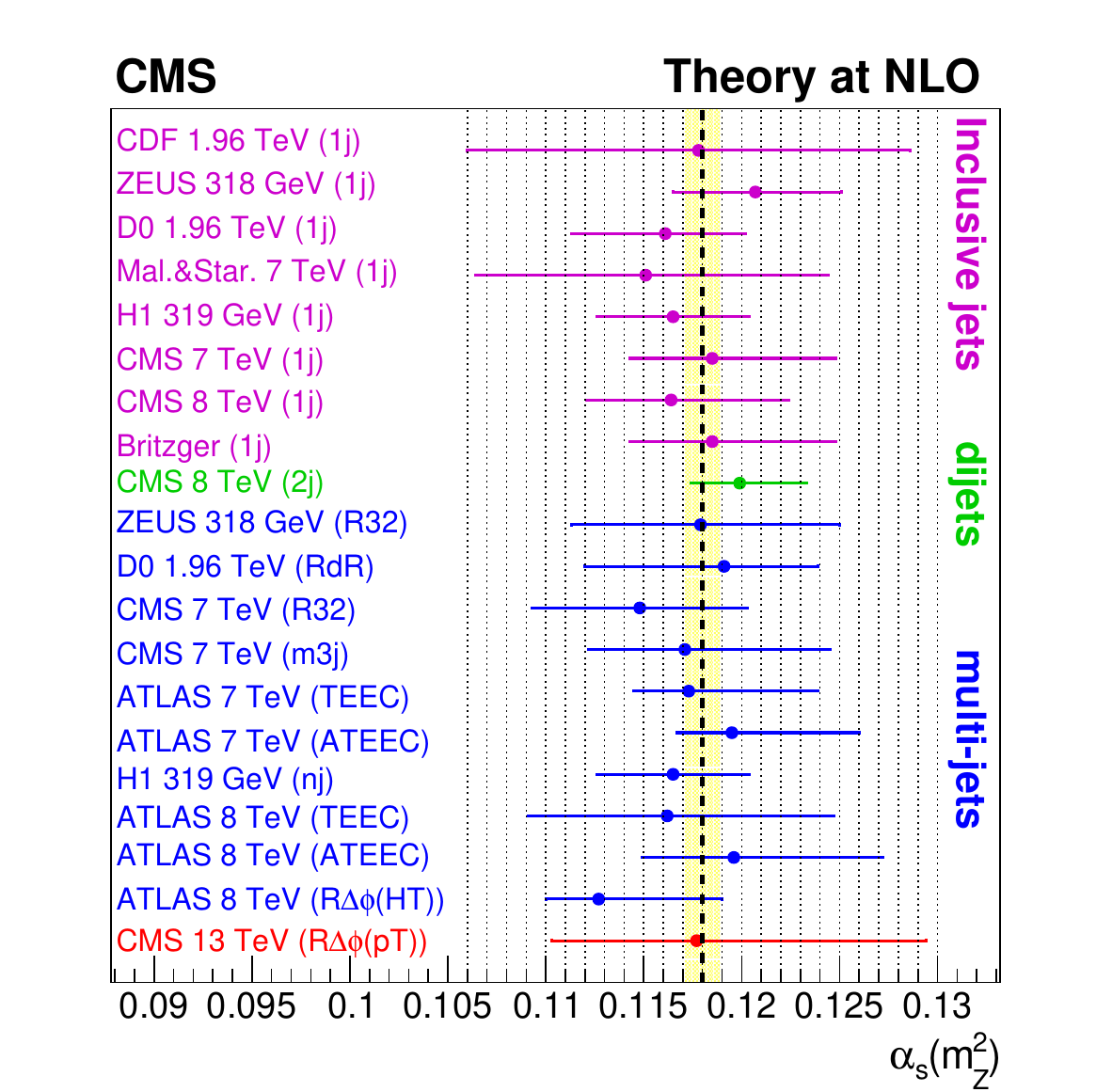}
  \caption{Determination of \alpsmz from the \rdphi ratio with the NNPDF3.1 PDF set (red), in comparison
    with previous NLO determinations of \alpsmz from inclusive jet (magenta), dijet (green), and multijet
   (blue) measurements. The horizontal error bars indicate the total uncertainty (experimental and theoretical).
    The world-average \alpsmz value is represented by the vertical dashed black line and its
    uncertainty by the yellow band.}
  \label{fig:AlphaSPDG}
\end{figure}

{\tolerance=800 The propagation of NP, PDF, and EW uncertainties is estimated separately by repeating the \chisq minimization procedure after
including the relevant terms in Eq.~(\ref{eq:covar}). For the evaluation of scale uncertainties, the \chisq comparison
between measurement and theoretical predictions is repeated for the six different combinations of \mur and \muf scales described in
Section~\ref{sec:TheoreticalPredictions}. The up/down scale uncertainties correspond to the difference between the
highest/lowest and the nominal \alpsmz values, respectively. All resulting \alpsmz values for the different PDF sets are
fully compatible among each other, as well as with the world average~\cite{PDG}. The spread of these \alpsmz values from the different PDF sets shown in Table~\ref{table:AlphasMz}, is used for the assignment of an additional uncertainty in the final \alpsmz result due to the PDF choice. This uncertainty is evaluated from the maximum difference among the \alpsmz values determined using the NNPDF3.1 NLO PDF set, and all the other \alpsmz values determined using the other PDF sets shown in Table~\ref{table:AlphasMz}.
The final result from the present analysis using the NNPDF3.1 NLO PDF set is: $\alpsmz = 0.1177_{-0.0068}^{+0.0114}~\,\text{(scale)} \pm 0.0013~\,\text{(exp)} \pm 0.0011~\,\text{(NP)} \pm 0.0010~\,\text{(PDF)} \pm 0.0003~\,\text{(EW)} \pm 0.0020~\,\text{(PDF choice)}$.
This result, in comparison with a selection of \alpsmz determinations at NLO accuracy obtained from inclusive
jet~\cite{bib:CDF,bib:ZEUS2,bib:D02,bib:ATLAS1,bib:H103, bib:7TeVas, bib:8TeVas, bib:Britzger}, dijet
\cite{bib:CMS8TeV3D}, and multijet~\cite{bib:ZEUS3, bib:D02, bib:R32, bib:3jetmass, bib:ATLAS7TeV,
  bib:H101, bib:H102, bib:H103, bib:ATLAS8TeVTEEC, bib:ATLAS8TeVazi} measurements is presented in Fig.~\ref{fig:AlphaSPDG}.\par}

\begin{table*}[htbp]
  \centering \renewcommand*{\arraystretch}{1.4}
  \topcaption{Results for \alpsmz, associated uncertainties, and goodness-of-fit per degree of freedom (\chisqndof), obtained from the measured \rdphi distribution compared with theoretical predictions using different NLO PDF sets.}
  \begin{tabular}{lccccccc}
    NLO PDF set & \alpsmz & Exp. & NP & PDF & EW  & Scale & \chisqndof \\
    \hline
    {ABMP16} & $0.1197$ & $0.0008$ & $0.0007$ &  $0.0007$  & $0.0002$ & $_{-0.0042}^{+0.0043}$   & $16/16$ \\
    {CT18} & $0.1159$ & $0.0013$ & $0.0009$ &  $0.0014$ & $0.0002$ & $_{-0.0067}^{+0.0099}$      & $19/16$ \\
    {MSHT20} & $0.1166$ & $0.0013$ & $0.0008$ &  $0.0010$ & $0.0003$ & $_{-0.0063}^{+0.0112}$    & $17/16$ \\
    {NNPDF3.1} & $0.1177$ & $0.0013$ & $0.0011$ &  $0.0010$ & $0.0003$ & $_{-0.0068}^{+0.0114}$   & $20/16$ \\
  \end{tabular}
  \label{table:AlphasMz}
\end{table*}

For the investigation of the running of the strong coupling, the fitted region of $\pt = 360$--3170\GeV (16 points) is
split into four \pt subregions: 360--700, 700--1190, 1190--1870, and 1870--3170\GeV (4 points each). The fitting procedure is
repeated and the \alpsmz and all the relevant uncertainties are determined in each subregion separately. The \alpsmz values from each subregion are
evolved to \alpsq, where $Q$ is chosen as the jet \pt and is calculated as a cross-section-weighted 
average $(\langle Q \rangle)$ for each subregion. This study is performed using the NNPDF3.1 NLO PDF set. The values of 
\alpsmz and the results for \alpsq evaluated at the respective $\langle Q \rangle$ for each fitted subregion are shown in 
Table~\ref{table:AlphasQ}.

\begin{figure*}[hbt!]
  \centering
  \includegraphics[width=0.98\textwidth]{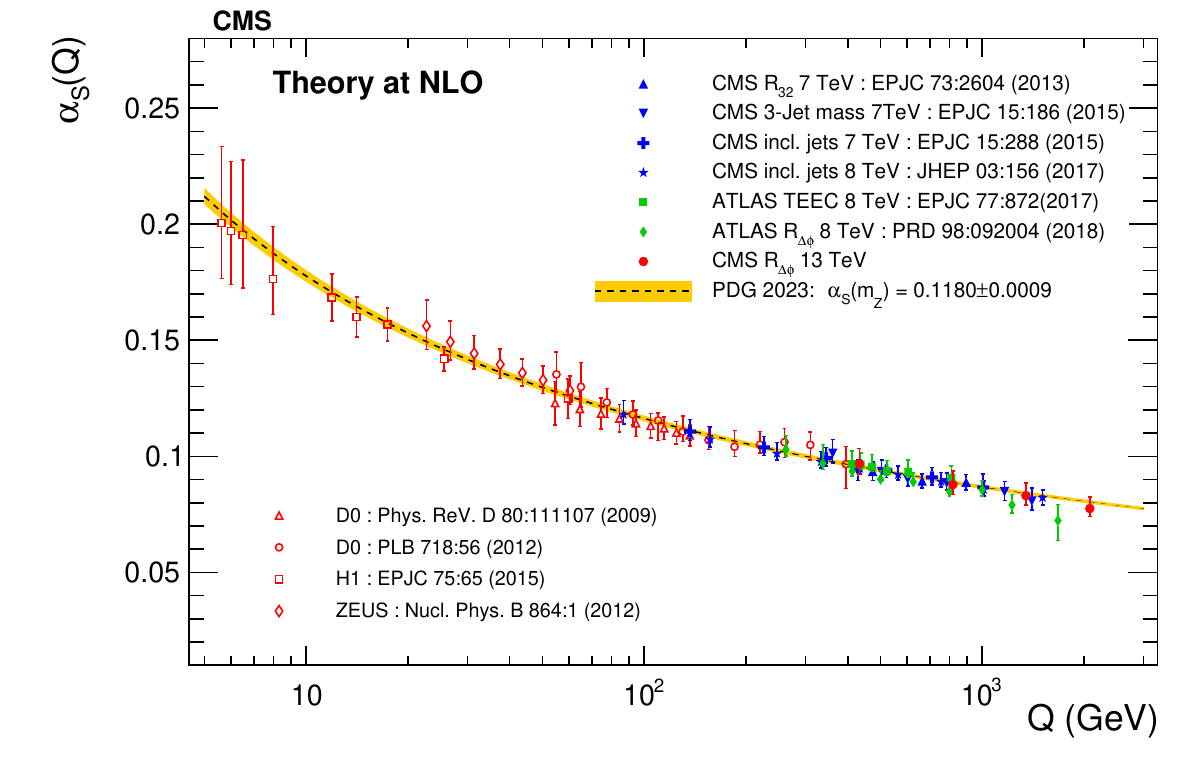}
  \caption{Running of the strong coupling \alpsq (dashed line) evolved using the current world-average value 
  	       $\alpsmz = 0.1180 \pm 0.0009$~\cite{PDG} together with its associated total uncertainty (yellow band). 
  	       The four new extractions from the present analysis (Table~\ref{table:AlphasQ}) are shown as filled red circles, 
  	       compared with results from the H1~\cite{bib:H101,bib:H102,bib:H103}, ZEUS~\cite{bib:ZEUS}, 
  	       D0~\cite{bib:D01,bib:D02}, CMS~\cite{bib:R32, bib:3jetmass, bib:7TeVas, bib:8TeVas}, and
  	       ATLAS~\cite{bib:ATLAS8TeVazi, bib:ATLAS8TeVTEEC} experiments. The vertical error bars indicate the total uncertainty
          (experimental and theoretical). All the experimental results shown in this figure
  	       are based on fixed-order predictions at NLO accuracy in pQCD.}
  \label{fig:Running}
\end{figure*}

Figure~\ref{fig:Running} shows the energy dependence predicted by the RGE (dashed line) using the current world-average value 
$\alpsmz = 0.1180 \pm 0.0009$~\cite{PDG} together with its associated total uncertainty (yellow band). The results from 
the \alpsq determinations in the four subregions presented in Table~\ref{table:AlphasQ} 
are also shown, along with \alpS values determined at lower scales by the H1~\cite{bib:H101,bib:H102,bib:H103}, ZEUS~\cite{bib:ZEUS},
D0~\cite{bib:D01,bib:D02}, CMS~\cite{bib:R32, bib:3jetmass, bib:7TeVas, bib:8TeVas}, and
ATLAS~\cite{bib:ATLAS8TeVazi,bib:ATLAS8TeVTEEC} Collaborations. All results reported in this study are consistent with the energy
dependence predicted by the RGE, and no deviation is observed from the expected behaviour up to $\sim2\TeV$.

\begin{table*}[htbp]
  \centering \renewcommand*{\arraystretch}{1.4}
  \topcaption{Values of \alpsmz and \alpsq determined in four different jet \pt fitting subregions corresponding to an average scale  $\langle Q \rangle$ over each \pt interval.}
  \begin{tabular}{cccc}
    \pt range ($\GeVns$) & \alpsmz & $\langle Q \rangle$ ($\GeVns$) & \alpsq \\
    \hline
    360--700 & $0.1177_{-0.0067}^{+0.0104}$ & 433.0 & $0.0967_{-0.0044}^{+0.0066}$     \\
    700--1190 & $0.1162_{-0.0073}^{+0.0108}$ & 819.0 & $0.0878_{-0.0042}^{+0.0060}$    \\
    1190--1870 & $0.1159_{-0.0077}^{+0.0112}$ & 1346.0 & $0.0830_{-0.0040}^{+0.0055}$  \\
    1870--3170 & $0.1118_{-0.0070}^{+0.0110}$ & 2081.0 & $0.0775_{-0.0034}^{+0.0051}$  \\
  \end{tabular}
  \label{table:AlphasQ}
\end{table*}

\section{Summary}
\label{sec:summary}

A measurement of the \rdphi ratio, sensitive to azimuthal correlations in multijet events, has been presented using proton-proton collision data collected by the CMS experiment at a centre-of-mass energy of 13\TeV and corresponding to an integrated luminosity of 134\fbinv. The experimental data are compared with predictions from Monte Carlo (MC) event generators, \PYTHIAEIGHT with tunes CUETP8M1 and CUETP8M2, \HERWIGpp with tune UE-EE-5-CTEQ6L1, and \POWHEG interfaced with each one of them. Deviations between data and MC predictions are observed in all cases, except for \PYTHIAEIGHT tune CUETP8M2, which gives a good overall description of the measurement.

The measurement is also compared with fixed-order perturbative quantum chromodynamics (pQCD) predictions at next-to-leading-order (NLO) accuracy using the \NLOJETPP package within the \fastNLO framework. Those predictions are extracted for four different NLO parton distribution function (PDF) sets, {ABMP16, CT18, MSHT20}, and {NNPDF3.1}. Corrections for nonperturbative (NP) effects are evaluated using all the aforementioned MC event generators, and are applied to the fixed-order predictions. The predictions are additionally corrected for electroweak (EW) effects that become important at large jet transverse momenta. Generally, the fixed-order predictions are in agreement with the experimental data in the phase space of this analysis, and they provide a good description of the measured \rdphi distribution for all PDF sets.

{\tolerance=2400 Based on a comparison of the measured \rdphi distribution and the theoretical predictions, the strong coupling at the scale of the \PZ boson mass is: $\alpsmz= 0.1177_{-0.0068}^{+0.0114}\,\text{(scale)}\pm 0.0013\,\text{(exp)}\pm 0.0011\,\text{(NP)}\pm 0.0010\,\text{(PDF)}\pm 0.0003\,\text{(EW)}\pm 0.0020\,\text{(PDF choice)} = 0.1177_{-0.0074}^{+0.0117}$, using calculations based on the NNPDF3.1 NLO PDF set. Alternative \alpsmz results obtained with other PDF sets are compatible among each other, as well as with the central result of this work, and with the current world average, $\alpsmz = 0.1180\pm0.0009$. The spread of the \alpsmz values obtained from different PDF sets is used for the assignment of the ``PDF choice'' uncertainty quoted in the final strong coupling constant derived here. The dominant uncertainty in this measurement originates from the scale dependence of the NLO pQCD predictions, and is expected to be significantly reduced with the future inclusion of fixed-order predictions at next-to-NLO accuracy.\par}

The evolution of the strong coupling as a function of the energy scale, \alpsq, has been tested up to $Q\approx2\TeV$, a higher scale than that probed in previous H1, ZEUS, D0, CMS, and ATLAS measurements. This test has been performed by choosing as energy scale $Q$ the average jet transverse momentum in the different intervals considered, and no deviation from the expected NLO pQCD running of the strong coupling is observed.

\begin{acknowledgments}

We thank Max Reyer, Marek Sch\"onherr and Steffen Schumann for providing the electroweak corrections for the \rdphi observable within the \SHERPA framework

We congratulate our colleagues in the CERN accelerator departments for the excellent performance of the LHC and thank the technical and administrative staffs at CERN and at other CMS institutes for their contributions to the success of the CMS effort. In addition, we gratefully acknowledge the computing centres and personnel of the Worldwide LHC Computing Grid and other centres for delivering so effectively the computing infrastructure essential to our analyses. Finally, we acknowledge the enduring support for the construction and operation of the LHC, the CMS detector, and the supporting computing infrastructure provided by the following funding agencies: SC (Armenia), BMBWF and FWF (Austria); FNRS and FWO (Belgium); CNPq, CAPES, FAPERJ, FAPERGS, and FAPESP (Brazil); MES and BNSF (Bulgaria); CERN; CAS, MoST, and NSFC (China); MINCIENCIAS (Colombia); MSES and CSF (Croatia); RIF (Cyprus); SENESCYT (Ecuador); ERC PRG, RVTT3 and MoER TK202 (Estonia); Academy of Finland, MEC, and HIP (Finland); CEA and CNRS/IN2P3 (France); SRNSF (Georgia); BMBF, DFG, and HGF (Germany); GSRI (Greece); NKFIH (Hungary); DAE and DST (India); IPM (Iran); SFI (Ireland); INFN (Italy); MSIP and NRF (Republic of Korea); MES (Latvia); LMTLT (Lithuania); MOE and UM (Malaysia); BUAP, CINVESTAV, CONACYT, LNS, SEP, and UASLP-FAI (Mexico); MOS (Montenegro); MBIE (New Zealand); PAEC (Pakistan); MES and NSC (Poland); FCT (Portugal); MESTD (Serbia); MCIN/AEI and PCTI (Spain); MOSTR (Sri Lanka); Swiss Funding Agencies (Switzerland); MST (Taipei); MHESI and NSTDA (Thailand); TUBITAK and TENMAK (Turkey); NASU (Ukraine); STFC (United Kingdom); DOE and NSF (USA).

{  \tolerance=1200
\hyphenation{Rachada-pisek} Individuals have received support from the Marie-Curie programme and the European Research Council and Horizon 2020 Grant, contract Nos.\ 675440, 724704, 752730, 758316, 765710, 824093, 101115353,101002207, and COST Action CA16108 (European Union); the Leventis Foundation; the Alfred P.\ Sloan Foundation; the Alexander von Humboldt Foundation; the Science Committee, project no. 22rl-037 (Armenia); the Belgian Federal Science Policy Office; the Fonds pour la Formation \`a la Recherche dans l'Industrie et dans l'Agriculture (FRIA-Belgium); the Agentschap voor Innovatie door Wetenschap en Technologie (IWT-Belgium); the F.R.S.-FNRS and FWO (Belgium) under the ``Excellence of Science -- EOS" -- be.h project n.\ 30820817; the Beijing Municipal Science \& Technology Commission, No. Z191100007219010 and Fundamental Research Funds for the Central Universities (China); the Ministry of Education, Youth and Sports (MEYS) of the Czech Republic; the Shota Rustaveli National Science Foundation, grant FR-22-985 (Georgia); the Deutsche Forschungsgemeinschaft (DFG), under Germany's Excellence Strategy -- EXC 2121 ``Quantum Universe" -- 390833306, and under project number 400140256 - GRK2497; the Hellenic Foundation for Research and Innovation (HFRI), Project Number 2288 (Greece); the Hungarian Academy of Sciences, the New National Excellence Program - \'UNKP, the NKFIH research grants K 131991, K 133046, K 138136, K 143460, K 143477, K 146913, K 146914, K 147048, 2020-2.2.1-ED-2021-00181, and TKP2021-NKTA-64 (Hungary); the Council of Science and Industrial Research, India; ICSC -- National Research Centre for High Performance Computing, Big Data and Quantum Computing and FAIR -- Future Artificial Intelligence Research, funded by the EU NexGeneration program (Italy); the Latvian Council of Science; the Ministry of Education and Science, project no. 2022/WK/14, and the National Science Center, contracts Opus 2021/41/B/ST2/01369 and 2021/43/B/ST2/01552 (Poland); the Funda\c{c}\~ao para a Ci\^encia e a Tecnologia, grant CEECIND/01334/2018 (Portugal); the National Priorities Research Program by Qatar National Research Fund; MCIN/AEI/10.13039/501100011033, ERDF ``a way of making Europe", and the Programa Estatal de Fomento de la Investigaci{\'o}n Cient{\'i}fica y T{\'e}cnica de Excelencia Mar\'{\i}a de Maeztu, grant MDM-2017-0765 and Programa Severo Ochoa del Principado de Asturias (Spain); the Chulalongkorn Academic into Its 2nd Century Project Advancement Project, and the National Science, Research and Innovation Fund via the Program Management Unit for Human Resources \& Institutional Development, Research and Innovation, grant B37G660013 (Thailand); the Kavli Foundation; the Nvidia Corporation; the SuperMicro Corporation; the Welch Foundation, contract C-1845; and the Weston Havens Foundation (USA).
\par}
\end{acknowledgments}

\bibliography{auto_generated}

\cleardoublepage \appendix\section{The CMS Collaboration \label{app:collab}}\begin{sloppypar}\hyphenpenalty=5000\widowpenalty=500\clubpenalty=5000
\cmsinstitute{Yerevan Physics Institute, Yerevan, Armenia}
{\tolerance=6000
A.~Hayrapetyan, A.~Tumasyan\cmsAuthorMark{1}\cmsorcid{0009-0000-0684-6742}
\par}
\cmsinstitute{Institut f\"{u}r Hochenergiephysik, Vienna, Austria}
{\tolerance=6000
W.~Adam\cmsorcid{0000-0001-9099-4341}, J.W.~Andrejkovic, T.~Bergauer\cmsorcid{0000-0002-5786-0293}, S.~Chatterjee\cmsorcid{0000-0003-2660-0349}, K.~Damanakis\cmsorcid{0000-0001-5389-2872}, M.~Dragicevic\cmsorcid{0000-0003-1967-6783}, P.S.~Hussain\cmsorcid{0000-0002-4825-5278}, M.~Jeitler\cmsAuthorMark{2}\cmsorcid{0000-0002-5141-9560}, N.~Krammer\cmsorcid{0000-0002-0548-0985}, A.~Li\cmsorcid{0000-0002-4547-116X}, D.~Liko\cmsorcid{0000-0002-3380-473X}, I.~Mikulec\cmsorcid{0000-0003-0385-2746}, J.~Schieck\cmsAuthorMark{2}\cmsorcid{0000-0002-1058-8093}, R.~Sch\"{o}fbeck\cmsorcid{0000-0002-2332-8784}, D.~Schwarz\cmsorcid{0000-0002-3821-7331}, M.~Sonawane\cmsorcid{0000-0003-0510-7010}, S.~Templ\cmsorcid{0000-0003-3137-5692}, W.~Waltenberger\cmsorcid{0000-0002-6215-7228}, C.-E.~Wulz\cmsAuthorMark{2}\cmsorcid{0000-0001-9226-5812}
% ...
\par}
\vskip\cmsinstskip
\dag:~Deceased\\
$^{1}$Also at Yerevan State University, Yerevan, Armenia\\
$^{2}$Also at TU Wien, Vienna, Austria\\
% ...
\end{sloppypar}
\end{document}